\documentclass[nofootinbib,amsmath,twocolumn,notitlepage,preprintnumbers]{revtex4-1}
\usepackage{multirow}
\usepackage{amssymb,esvect,amsmath,graphicx,latexsym,amsthm,slashed,eso-pic}

\usepackage{siunitx}
\usepackage{hyperref}
\usepackage{placeins}

\newcommand{\beq}{\begin{equation}} \newcommand{\eeq}{\end{equation}}
\newcommand{\bea}{\begin{eqnarray}} \newcommand{\eea}{\end{eqnarray}}

\newcommand{\Ts}{T_\mathrm{dom}}

\newcommand{\aap}{Astron. Astrophys.}

\newcommand{\jcap}{JCAP}

\newcommand{\mnras}{Mon. Not. R. Astron. Soc.}

\newcommand{\lsim}{\mathrel{\hbox{\rlap{\lower.55ex\hbox{$\sim$}} \kern-.3em \raise.4ex \hbox{$<$}}}}
\newcommand{\gsim}{\mathrel{\hbox{\rlap{\lower.55ex\hbox{$\sim$}} \kern-.3em \raise.4ex \hbox{$>$}}}}
\newcommand{\drm}{\mathrm{d}}

\newcommand{\krh}{k_\mathrm{RH}}

\newcommand{\mdec}{M_\mathrm{RH}}

\newcommand{\delcoll}{\delta_c}

\newcommand{\sigv}{\langle \sigma v \rangle}

\newcommand{\Trh}{T_\mathrm{RH}}
\newcommand{\mpl}{m_\mathrm{Pl}}
\newcommand{\mdm}{m_\chi}

\newcommand{\kcut}{k_\mathrm{cut}}

\def\lsim{\mathrel{\raise.3ex\hbox{$<$\kern-.75em\lower1ex\hbox{$\sim$}}}}
\def\gsim{\mathrel{\raise.3ex\hbox{$>$\kern-.75em\lower1ex\hbox{$\sim$}}}}

\newcommand{\be}{\begin{eqnarray}}
\newcommand{\ee}{\end{eqnarray}}

\newcommand{\benum}{\begin{enumerate}}
\newcommand{\eenum}{\end{enumerate}}
\newcommand{\bi}{\begin{itemize}}
\newcommand{\ei}{\end{itemize}}

\begin{document}

\preprint{FERMILAB-PUB-19-249-A}

\title{Annihilation Signatures of Hidden Sector Dark Matter Within Early-Forming Microhalos}

\author{Carlos Blanco$^{a,b}$}
\thanks{ORCID: http://orcid.org/0000-0001-8971-834X}

\author{M.~Sten Delos$^c$}
\thanks{ORCID: http://orcid.org/0000-0003-3808-5321}

\author{Adrienne L.~Erickcek$^{c}$}
\thanks{ORCID: http://orcid.org/0000-0002-0901-3591}

\author{Dan Hooper$^{b,d,e}$}
\thanks{ORCID: http://orcid.org/0000-0001-8837-4127}

\affiliation{$^a$University of Chicago, Department of Physics, Chicago, IL 60637, USA}
\affiliation{$^b$University of Chicago, Kavli Institute for Cosmological Physics, Chicago, IL 60637, USA}
\affiliation{$^c$Department of Physics and Astronomy, University of North Carolina at Chapel Hill, Phillips Hall CB 3255, Chapel Hill, NC 27599, USA}
\affiliation{$^d$Fermi National Accelerator Laboratory, Theoretical Astrophysics Group, Batavia, IL 60510, USA}
\affiliation{$^e$University of Chicago, Department of Astronomy and Astrophysics, Chicago, IL 60637, USA}

\date{\today}

\begin{abstract}

If the dark matter is part of a hidden sector with only very feeble couplings to the Standard Model, the lightest particle in the hidden sector will generically be long-lived and could come to dominate the energy density of the universe prior to the onset of nucleosynthesis. During this early matter-dominated era, density perturbations will grow more quickly than otherwise predicted, leading to a large abundance of sub-earth-mass dark matter microhalos.   Since the dark matter does not couple directly to the Standard Model, the minimum halo mass is much smaller than expected for weakly interacting dark matter, and the smallest halos could form during the radiation-dominated era. In this paper, we calculate the evolution of density perturbations within the context of such hidden sector models and use a series of $N$-body simulations to determine the outcome of nonlinear collapse during radiation domination. The resulting microhalos are extremely dense, which leads to very high rates of dark matter annihilation and to large indirect detection signals that resemble those ordinarily predicted for decaying dark matter. We find that the Fermi Collaboration's measurement of the high-latitude gamma-ray background rules out a wide range of parameter space within this class of models.  The scenarios that are most difficult to constrain are those that feature a very long early matter-dominated era; if microhalos form prior to the decay of the unstable hidden sector matter, the destruction of these microhalos effectively heats the dark matter, suppressing the later formation of microhalos.

\end{abstract}

\maketitle

\section{Introduction}

The null results of direct detection experiments~\cite{Aprile:2018dbl,Akerib:2016vxi,Cui:2017nnn}, as well as searches for new physics at the Large Hadron Collider~\cite{Sirunyan:2018dub,Sirunyan:2018xlo,Sirunyan:2018fpy,Sirunyan:2018wcm,Sirunyan:2018gka,Sirunyan:2017leh,Aaboud:2018xdl,Aaboud:2017phn,Aaboud:2017rzf,Aaboud:2017bja,Aaboud:2017yqz}, have provided motivation to consider scenarios in which the dark matter does not couple directly to the particle content of the Standard Model (SM). In particular, it is plausible that the dark matter particles do not couple to the SM, but instead annihilate into one or more unstable species that reside within a hidden sector~\cite{Pospelov:2007mp,ArkaniHamed:2008qn,Abdullah:2014lla,Berlin:2014pya,Martin:2014sxa,Hooper:2012cw}. If these annihilation products are short-lived, the evolution of the early universe will closely resemble that found in the standard case of thermal freeze-out. If they are long-lived, however, such hidden sector annihilation products could become non-relativistic and come to dominate the energy density of the universe before they decay. In such a scenario, the decays of these particles would reheat the universe, producing a new bath of SM radiation and diluting the abundances of all previously formed relics, including that of the dark matter~\cite{BHK16a,BHK16b} (see also Refs.~\cite{Dror:2016rxc,Dror:2017gjq}).

Long lifetimes are easily realized in hidden sector models, in particular if the decaying particle is the lightest hidden sector state. Furthermore, if the coupling between the hidden sector and the SM is very small, these two sectors will remain decoupled from one another throughout thermal history. Such a picture is a relatively generic feature of models in which the dark matter freezes out through annihilations within a highly decoupled hidden sector.  While a variety of simple and well-motivated hidden sector dark matter scenarios have been proposed, it has often proven difficult to constrain or otherwise test these models. In particular, the feeble couplings between the hidden and SM sectors makes the prospects for dark matter searches at colliders and direct detection experiments very bleak. 

The prospects for indirect detection in hidden sector dark matter models are less uniformly negative, but they depend strongly on the thermal history of the hidden sector. In scenarios in which the unstable hidden sector particles decay while they are still relativistic, the dark matter freezes out during radiation domination, requiring an annihilation cross section near $\sigma v \simeq 2\times 10^{-26}$ cm$^3/$s in order to yield the measured dark matter abundance~\cite{Steigman:2012nb}. In this case, measurements by gamma-ray telescopes and cosmic-ray detectors can constrain a wide range of such models, for dark matter masses up to $\sim$\,$100$ GeV~\cite{Fermi-LAT:2016uux,Cholis:2019ejx,Bergstrom:2013jra}. If the hidden and SM sectors are more weakly coupled, however, the eventual out-of-equilibrium decay of the hidden sector would dilute the dark matter, requiring a smaller dark matter annihilation cross section to obtain the measured abundance and suppressing the predicted indirect detection signal. 
    
In this article, we study the growth of small-scale structure in hidden sector dark matter scenarios and discuss the impact of such structure on the prospects for indirect detection. During an early matter-dominated era (EMDE), matter density perturbations grow linearly, in contrast to the logarithmic growth that occurs during standard radiation domination~\cite{ES11, BR14, FOW14, E15}.  This rapid growth enhances small-scale inhomogeneity and can trigger the formation of numerous sub-earth-mass microhalos long before structure would form in the absence of an EMDE.  The microhalo abundance at early times is primarily limited by the temperature of the dark matter particles: the velocity dispersion of dark matter erases perturbations below the free streaming scale.  For weakly interacting dark matter, the first microhalos generally form at redshifts less than 1000 \cite{ESW16}, but microhalos can form much earlier in hidden sector scenarios because the dark matter particles are not in kinetic equilibrium with the SM and are therefore colder.  

In many scenarios we consider, perturbation growth during the EMDE leads to structure formation long before the universe becomes matter dominated, and we develop a theory of gravitational collapse during radiation domination to predict the properties of these microhalos.  Although the presence of these microhalos would not have any discernible impact on the observable structure of our universe, the extraordinarily high dark matter annihilation rates within their dense cores leads to detectable fluxes of gamma rays and cosmic rays.  We also point out that the angular distribution of this gamma-ray signal is distinct from that predicted from ordinary annihilating dark matter, instead mimicking that normally associated with decaying dark matter particles.  We show that the annihilation signal can be described by an effective dark matter lifetime, and we use the Fermi Collaboration's measurement of the high-latitude gamma-ray background~\cite{Ackermann:2014usa} and resulting constraints on dark matter decay \cite{Blanco:2018esa} to constrain these scenarios.  

The lack of dark matter free streaming also opens up the possibility that even smaller microhalos could form during the EMDE if it lasts long enough for perturbations to collapse.  We show that these halos would be destroyed when most of their content decays into relativistic particles.  The freed dark matter particles would be released with much larger velocities than they had before their infall into a halo, and their subsequent free streaming suppresses the formation of microhalos after the EMDE.  This gravitational heating effectively erases the signatures of the EMDE, making these scenarios difficult to constrain.

In the following section, we describe the thermal history of the early universe within the context of a vector portal hidden sector model, including the freeze-out of the dark matter, the subsequent EMDE, and the reheating of the SM bath through the decay of the unstable hidden sector matter. We then discuss in Secs.~\ref{during} and ~\ref{after} the evolution of density perturbations during and after the EMDE, leading to the formation of a large population of sub-earth-mass microhalos. In Sec.~\ref{heating} we discuss how structure formation during the EMDE suppresses microhalo formation after the EMDE.  Our main results are presented in Sec.~\ref{results}, in which we calculate the dark matter's annihilation rate and place constraints on this class of models from measurements of the high-latitude gamma-ray background. Finally, we summarize our results and conclusions in Sec.~\ref{summary}.

\section{Vector Portal Dark Matter}

A variety of hidden sector models have been described in the literature, including those in which the hidden sector couples to the SM through the renormalizable interactions known as the vector portal ($B^{\mu \nu}$)~\cite{Pospelov:2007mp,Krolikowski:2008qa}, the Higgs portal ($H^{\dagger} H$)~\cite{Pospelov:2007mp,Burgess:2000yq,Davoudiasl:2004be,Bird:2006jd,Kim:2006af,Finkbeiner:2007kk,DEramo:2007anh,MarchRussell:2008yu}, and the lepton portal ($H^{\dagger} L$)~\cite{Pospelov:2007mp,Bai:2014osa}. For concreteness, we will focus in this study on a vector portal model, although most of our results apply to each of these cases. We take our dark matter candidate to be a stable Dirac fermion, $X$, that couples to a gauge boson, $Z'$:
\begin{eqnarray}
\mathcal{L} \supset -\frac{\epsilon}{2} B^{\mu \nu} Z'_{\mu \nu} + g_{X} Z'_{\mu} \bar{X} \gamma^{\mu} X.
\end{eqnarray}
The first term in this expression describes the kinetic mixing between the $Z'$ and SM $Z$/photon~\cite{Holdom:1985ag}. Through this kinetic mixing, the $Z'$ acquires the following couplings to SM fermions:
\begin{equation}
g_{f_{R,L}} = \epsilon \bigg(\frac{m^2_{Z'} g_Y Y_{f_{R,L}}-m^2_Z g \sin \theta_W \cos \theta_W Q_f}{m^2_Z-m^2_{Z'}}\bigg), 
\end{equation}
where $\theta_W$ is the weak mixing angle, $m_Z$ is the mass of the SM $Z$ and $g_Y$ and $g$ are the $U(1)_Y$ and $SU(2)_L$ gauge couplings, respectively. 
As a result of these couplings, the $Z'$ decays to SM fermions with the following width: 
\begin{equation}
\label{Gamma}
\Gamma_{Z'} = \sum_f \frac{n_c m_{Z'} \beta_f}{12 \pi} \bigg[g^2_{fv} \bigg(1+\frac{2 m_f^2}{m^2_{Z'}}\bigg) +g^2_{fa} \bigg(1-\frac{4m_f^2}{m^2_{Z'}}\bigg)\bigg],
\end{equation}
where $n_c$ is the number of colors of the final state fermions and $\beta_f \equiv \sqrt{1-4m^2_f/m^2_{Z'}}$. For $m_{Z'} \gg m_Z, m_f$, this reduces to a lifetime given by
\begin{equation}
\tau_{Z'} \approx 4 \times 10^{-6} \, {\rm s} \, \times \bigg(\frac{10^{-10}}{\epsilon}\bigg)^2 \bigg(\frac{{\rm TeV}}{m_{Z'}}\bigg).
\end{equation}

Dark matter annihilation proceeds through $t$-channel $X$ exchange in this model to produce a $Z' Z'$ pair, with a cross section that is given by
\begin{equation}
\frac{1}{2}\sigma v (XX\rightarrow Z'Z') \approx \frac{\pi \alpha^2_X}{m^2_X} \bigg(1-\frac{7v^2}{12}\bigg),
\end{equation}
where $v$ is the relative velocity between the annihilating particles and $\alpha_X \equiv g^2_X/4\pi$.

For $\epsilon\gsim 3\times 10^{-8} \times (T/{\rm GeV})^{1/2} \, (g_{\star}/75)^{1/4}$, where $g_{\star}$ is the effective number of degrees-of-freedom at temperature $T$, interactions of the $Z'$ will maintain equilibrium in the early universe between the hidden sector and the SM (see Appendix 7 of Ref.~\cite{BHK16a}). For smaller values of $\epsilon$, however, the hidden sector will be entirely decoupled from the SM bath, allowing the two sectors to evolve independently and maintain their own temperatures. We define $\xi \equiv T_h/T$ as the ratio of the temperatures of the hidden and SM sectors. Following from entropy conservation, the value of $\xi$ evolves as
\begin{equation}
\xi = \xi_{\rm inf} \bigg(\frac{g^h_{\star, {\rm inf}}}{g^{h}_{\star}}\bigg)^{1/3} \bigg(\frac{g_{\star}}{g_{\star,{\rm inf}}}\bigg)^{1/3},
\end{equation}
where $g_{\star}$ and $g^h_{\star}$ are the numbers of effective relativistic degrees-of-freedom in the SM and hidden sectors, respectively, and the ``inf'' subscript denotes the initial value after inflation.  

If the $Z'$ population remains close to its equilibrium value during dark matter freeze-out (for details, see Ref.~\cite{BHK16b}) and decays before becoming non-relativistic, the resulting dark matter thermal relic abundance is given by
\begin{eqnarray}
\label{relic1}
  \hspace{-0.1cm} \Omega_X h^2 &\approx& 8.5 \times 10^{-11} ~ \frac{x_f \sqrt{g_\star^\text{eff}}}{g_\star} ~ \left( \frac{a+3 \xi b / x_f}{\text{GeV}^{-2}} \right)^{-1} ~~ \\
   && \hspace{-0.3cm} \approx   0.11  \times \bigg(\frac{x_f}{20}\bigg)  \bigg(\frac{0.03}{\alpha_X}\bigg)^2  \bigg(\frac{m_X}{{\rm TeV}}\bigg)^2
\bigg(\frac{\sqrt{g^{\rm eff}_{\star}}/g_{\star}}{0.1} \!  \bigg),~~~~ \label{relic2} \nonumber
\end{eqnarray}
where $a$ and $b$ are terms in the expansion of the dark matter annihilation cross section, $\sigma v/2 \approx a +b v^2 + \mathcal{O}(v^4)$, and $g^{\rm eff}_{\star} \equiv g_{\star} + g^h_{\star} \, \xi^4$ at freeze-out. $x_f$ is defined as the mass of $X$ divided by the SM temperature at freeze-out, and is found to have a value of $\sim 20 \times \xi$ over a wide range of parameters.

The dark matter relic abundance can be very different from that described in the above expression if the two sectors are not maintained in equilibrium in the early universe. In this case, the energy density of the $Z'$ population after they have become non-relativistic and prior to their decays is given by
%
%
\begin{equation}
\rho_{Z'} = \frac{13\zeta(3) g_{\star} m_{Z'} \xi_{\rm inf}^3 T^3}{2\pi^2 g_{\star, {\rm inf}}} ,
\end{equation}
where $T$ is the temperature of the SM bath and $g_{\star,\mathrm{inf}} \simeq 106$. Included in this expression is factor of $13/6$ which is equal to one plus the ratio of the internal degrees of freedom associated with the $X$ and $Z'$ (for details, see Ref.~\cite{BHK16b}). The EMDE begins when this density exceeds that of the SM radiation bath ($\rho_{Z'} > \rho_{\rm SM} = \pi^2g_{\star} T^4/30$), which occurs at the following temperature:
\begin{eqnarray}
T_{\rm dom} &=& \frac{195\zeta(3) m_{Z'}  \xi^{3}_{\rm inf}}{\pi^4 g_{\star,\mathrm{inf}}} , \nonumber \\
&\approx & 22 \, {\rm GeV} \times \xi_{\rm inf}^3 \, \bigg(\frac{m_{Z'}}{{\rm TeV}}\bigg).
\end{eqnarray}

The EMDE ends when $Z'$ particles decay into SM particles, and this decay rate defines the reheat temperature:\footnote{Note that this definition of $T_{\rm RH}$ is equal to $1.22 \times T_{f}$ as used in Ref.~\cite{BHK16a}.}
\beq
\Gamma_{Z'} \equiv  \sqrt{\frac{8\pi^3 g_\star(\Trh)}{90}}\frac{\Trh^2}{\mpl},
\label{Trhdef}
\eeq
which roughly corresponds to the temperature at the onset of radiation domination following the EMDE.  Setting this equal to the decay rate given in Eq.~(\ref{Gamma}), we find that the SM bath is heated to the following temperature:
\begin{equation}
T_{\rm RH} \approx 620 \, {\rm MeV} \bigg(\frac{\epsilon}{10^{-10}}\bigg)\bigg(\frac{m_{Z'}}{{\rm TeV}}\bigg)^{1/2} \bigg(\frac{10}{g_{\star}}\bigg)^{1/4},
\label{TRH}
\end{equation}
and
\begin{equation}
\frac{T_{\rm dom}}{T_{\rm RH}} \simeq 36 \, \times \xi_{\rm inf}^3 \, \bigg(\frac{10^{-10}}{\epsilon}\bigg) \bigg(\frac{m_{Z'}}{{\rm TeV}}\bigg)^{1/2} \bigg(\frac{g_{\star}}{10}\bigg)^{1/4}.
\label{TdomTRH}
\end{equation}

The decays of the $Z'$ population dilute the relic abundance of dark matter by a factor of $T_{\rm dom}/T_{\rm RH}$, leading to the following final dark matter relic abundance~\cite{BHK16a,BHK16b}:
\begin{eqnarray}
~\Omega_X h^2 &\approx & \frac{0.11}{\xi_{\rm inf}^3} \,         \bigg(\frac{\epsilon}{10^{-10}}\bigg) \, \bigg(\frac{0.05}{\alpha_X}\bigg)^2 \, \bigg(\frac{m_X}{10\,{\rm TeV}}\bigg)^2 \, \bigg(\frac{{\rm TeV}}{m_{Z'}}\bigg)^{1/2}   \nonumber  \\
    &\times&    \bigg(\frac{x_f}{20}\bigg)    \bigg(\frac{\sqrt{g^{\rm eff}_{\star}}/g_{\star}}{0.1}\bigg) \,
  \bigg(  \frac{10}{\langle g^{1/3}_{\star} \rangle^{3}} \bigg)^{1/4},
\end{eqnarray}
where $\langle g_{\star} \rangle$ denotes the time-averaged value over the period of decay~\cite{Kolb:1990vq}.

%
%
%

\section{Growth of Perturbations during an Early Matter-Dominated Era}
\label{during}

In this section, we study the growth of density perturbations during an era in the early universe in which the total energy density is dominated by non-relativistic hidden sector particles. Within this context, we consider a universe that contains three components: the dark matter, $X$, the unstable but relatively long-lived $Z'$, and the relativistic SM bath.  We assume that the universe was initially dominated by SM particles and that, during this epoch, the dark matter thermally decoupled from the $Z'$ population while the $Z'$ particles were still relativistic.  The $Z'$ population then became nonrelativistic and came to dominate the universe, leading to an EMDE.   The duration of the EMDE and the subsequent reheating temperature are determined by the properties of the $Z'$ particles, as described in the previous section.  

The evolution of density perturbations during an EMDE has been studied extensively \cite{ES11, BR14, FOW14, E15}.  In this study, we use the suite of linear perturbation equations in Newtonian gauge derived in Ref.~\cite{ES11} for a pressureless scalar field that decays into relativistic particles and dark matter. In our case, however, the pressureless scalar field is replaced by the $Z'$ particles, and these particles do not decay into dark matter particles.  We also start our perturbation evolution prior to the EMDE, and we do not assume that all of the radiation present is sourced by the decay of the $Z'$ population. Our initial conditions, therefore, differ from those used in Ref.~\cite{ES11}. In particular, assuming that the $Z'$ population is nonrelativistic, superhorizon adiabatic initial conditions during radiation domination demand that the fractional density perturbation in the SM particles, $\delta_R \equiv (\rho_R-\bar{\rho}_R)/\bar{\rho}_R$, is related to the curvature perturbation ($\delta_R = 2 \Phi$, where $\Phi$ is defined as in Ref.~\cite{ES11}) and to the fractional perturbations in the densities of $X$ and $Z'$ particles: $\delta_X = \delta_{Z'} = (3/4) \delta_R$.  While the perturbation mode is outside the horizon, the comoving divergence of each fluid's physical velocity, $\theta \equiv a\, \partial_i (dx^i/dt)$, is equal to $-(\Phi/2)k^2/(aH)^2$, where $k$ is the wave number of the mode and $H$ is the Hubble parameter.

\begin{figure}
 \resizebox{3.4in}{!}
 {
      \includegraphics{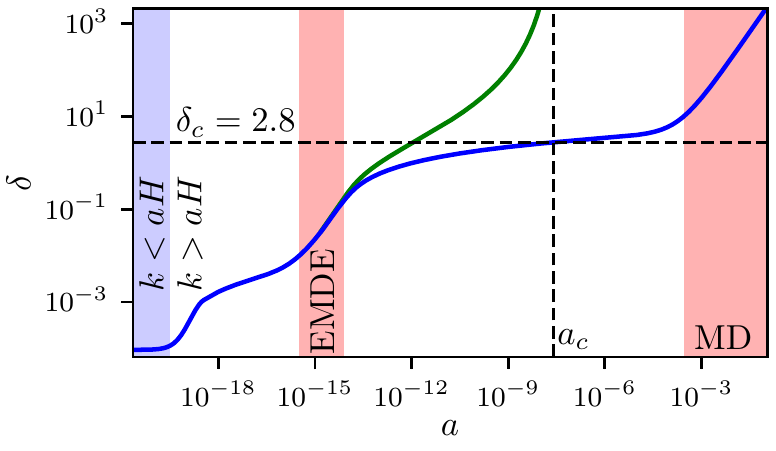}
 }
\caption{The evolution of the fractional density perturbations in dark matter for a mode with $k  = 1.5\times 10^7$ pc$^{-1}$ that enters the horizon prior to an EMDE with $T_\mathrm{RH} = 16$ GeV 
and $\Ts = 233$ GeV.  Upon entering the horizon ($k=aH$), $\delta_X$ grows logarithmically with the scale factor, $a$.  It transitions to linear growth during the EMDE and then returns to logarithmic growth after reheating.  This mode goes nonlinear during the radiation-dominated era, and the top (green) curve shows its evolution according to spherical collapse theory.  The bottom (blue) curve shows the continued linear-theory evolution; $\delta$ resumes linear growth when the universe becomes matter dominated (MD).}
\label{Fig:growth}
\end{figure}

After the $Z'$ population has become nonrelativistic and prior to the EMDE, $\delta_X$ and $\delta_{Z'}$ each grow logarithmically with the scale factor after the perturbation mode enters the horizon (when $k=aH$).  This growth is triggered by the gravitational pull toward overdense regions at horizon entry, but the decay of subhorizon gravitational potential perturbations during radiation domination implies that these forces quickly fade.  The subsequent logarithmic growth in $\delta$ shown in Fig.~\ref{Fig:growth} is a consequence of the continued drift of the dark matter particles toward regions that were initially overdense.\footnote{In the absence of gravitational forces, comoving velocities decay as $a^{-2}$, so the comoving displacement of a particle, $\vec s$, is proportional to ${\int a^{-2} dt = \int a^{-2} da/(aH) \propto \int da/a}$, since $H \propto a^{-2}$ during radiation domination.  Thus, $\vec s\propto \ln a$, and since $\delta = -\nabla \cdot \vec s$ at linear order, $\delta \propto \ln a$.}  
When the $Z'$ population comes to dominate the universe at the start of the EMDE, the gravitational forces reappear and pull both the $X$ and $Z'$ particles toward the regions that have more $Z'$ particles than average, prompting both $\delta_X$ and $\delta_{Z'}$ to transition to linear growth.  At the end of the EMDE, $\rho_{Z'} \simeq 0$, and so the perturbations in $\rho_{Z'}$ are no longer relevant.  As shown in Fig.~\ref{Fig:growth}, $\delta_X$ transitions to logarithmic growth after the EMDE and continues to grow logarithmically with the scale factor until it either approaches unity, which signifies the breakdown of linear theory, or the universe becomes matter dominated, at which point $\delta_X$ resumes linear growth.  Fig.~\ref{Fig:growth} also illustrates that the perturbation mode continues to grow rapidly for several e-folds of expansion after the EMDE ends, which implies that even a short EMDE can significantly enhance the amplitude of dark matter density fluctuations.  This post-EMDE growth reflects the fact that the particles were accelerated toward the overdense regions during the EMDE, leading to a faster drift toward those regions after the EMDE ends.

Our suite of perturbation equations does not include baryons, which means that it does not accurately model the evolution of $\delta_X$ after the baryons become nonrelativistic.  Baryons do not participate in structure growth on the small scales that enter the horizon prior to reheating \citep[e.g.][]{Bertschinger06}, and the growth of $\delta_X$ is suppressed as a result.  To account for this suppression, we solve our perturbation equations until logarithmic growth is established after reheating ($a\simeq 1000 \, a_\mathrm{RH}$, where $a_{\rm RH}$ is the scale factor at the time of reheating).  We then match this evolution to the solutions to the Meszaros equation provided in Ref.~\cite{HS96} that apply while the baryons are coupled to the photons and therefore do not participate in structure growth.  We use these solutions to evolve $\delta_X$ past $a=1000 \, a_\mathrm{RH}$. 

Perturbation modes that enter the horizon while the $Z'$ population is still relativistic are not accurately described by our suite of perturbation equations.  While the $Z'$ particles are relativistic, their pressure will prevent perturbation growth, and $\delta_{Z'}$ will oscillate.  Since the $X$ particles remain kinetically coupled to the $Z'$ particles, they will inherit these oscillations, but the oscillations in $\delta_X$ will be damped \cite{Bertschinger06}.  We account for the suppression of perturbations due to these damped oscillations by introducing an exponential cut-off in the matter power spectrum: $P(k) \propto \exp[-k^2/k_{Z'}^2]$, where $k_{Z'}$ is the wavenumber of the mode that enters the horizon when the temperature of the hidden sector equals $m_{Z'}$. In Sec.~\ref{results}, we explore how altering this cut-off scale impacts the microhalo population and the resulting gamma-ray constraints.

\section{Structure Formation Following an Early Matter-Dominated Era}
\label{after}

The perturbation mode whose evolution is shown in Fig.~\ref{Fig:growth} enters the nonlinear regime while the universe is radiation dominated.  To track the evolution of this perturbation, we generalize the spherical collapse model \cite{Peebles80, PS93} so that we can apply it to cosmologies that include an EMDE.  The evolution of a spherical shell of matter that surrounds an overdense region that has an average matter density that is $(1+\delta)$ times the background density is governed by the second Friedmann equation: its physical radius, $R$, obeys
\begin{equation} 
\frac{1}{R}\frac{d^2R}{dt^2} = -\frac{4\pi G}{3}\left(\rho_X + \rho_{Z'} + 2 \bar{\rho}_R  - 2\bar{\rho}_\Lambda\right),
\label{Rdoubledot}
\end{equation}
where $\bar{\rho}_\Lambda$ and $\bar{\rho}_R$ are the energy densities associated with the cosmological constant and the relativistic SM particles respectively.\footnote{For simplicity, we do not include changes to the radiation's composition in our spherical collapse formulation.  To account for such variations, the $2\bar\rho_R$ term should be replaced with $\bar\rho_R+3\bar P_R$, where $\bar P_R$ is the radiation pressure.}  These densities are unperturbed; we neglect the perturbations in the relativistic particles because these perturbations are suppressed after reheating \cite{ES11, BR14, FOW14}.  The average density of $X$ and $Z'$ particles inside the shell is given by
\begin{equation}
\rho_X+\rho_{Z'} = \left[\bar{\rho}_{X,i}+\left(\frac{\bar{\rho}_{Z'} a^3}{\bar{\rho}_{Z',i} a_i^3} \right)\bar{\rho}_{Z',i}\right](1+\delta_i)\left(\frac{R_i}{R}\right)^3,
\eeq
where a subscript $i$ indicates that the quantity is evaluated at some initial time, and $\bar{\rho}$ refers to the homogenous background density.  This equation accounts for the decay of the $Z'$ particles by multiplying the mass of $Z'$ particles enclosed within the shell by the change in their comoving number density.  The initial time is chosen to be during the EMDE such that $\delta_i \ll 1$ according to linear theory.  With this selection, the Zel'dovich approximation \cite{Zeldovich70} provides an initial condition for $dR/dt$:
\beq
\frac{dR}{dt} = H_i R_i \left(1-\frac{\delta_i}3\right).
\eeq
Since the shells do not cross, any value of $R_i$ gives the same evolution for the density within the shell: 
\beq
1+\delta_{\mathrm{SC}} = (1+\delta_i)\left(\frac{a}{a_i}\right)^3\left(\frac{R_i}{R}\right)^3.
\eeq

Fig.~\ref{Fig:growth} shows the evolution of $\delta_\mathrm{SC}$ that is obtained by solving Eq.~(\ref{Rdoubledot}) for $R(a)$ with these initial conditions.  When the value of $\delta_\mathrm{SC}$ is much less than one, the spherical collapse model matches the linear theory, but $\delta_\mathrm{SC}$ increases faster than the linear solution as $\delta$ approaches unity and eventually diverges at a finite time.  This divergence indicates that the spherical shell has collapsed to $R=0$, and we denote value of the scale factor at this time as $a_c$.  We also define $\delta_c \equiv \delta(a_c)$ as predicted by linear theory.  For the perturbation mode shown in Fig.~\ref{Fig:growth}, $\delta_c = 2.8$, which is significantly greater than its value during matter domination $(\delta_c = 1.686)$.  Thus Fig.~\ref{Fig:growth} illustrates that gravitational collapse does occur while the universe is radiation dominated, but it occurs later than it would in a matter-dominated universe.  By considering different values for $\delta_i$, corresponding to different perturbation wavenumbers, we can determine how $\delta_c$ varies with $a_c$: it equals 1.68 during the EMDE and then increases up to $\delta_c \simeq 2.5$ before falling back to $\sim\!\!1.68$ at $a_c\gsim a_\mathrm{eq}$. While the precise shape of $\delta_c(a_c)$ depends on $T_\mathrm{RH}$, $\delta_c$ for $a > 10^{-6}$ is nearly the same for all $T_\mathrm{RH} >2$ MeV.  In Appendix~\ref{fitting}, we plot and provide a fitting function for $\delta_c(a_c, T_\mathrm{RH})$.

Since dark matter particles do not experience any gravitational forces while the universe is radiation dominated, the collapse of a spherical shell does not necessarily indicate that a dark matter halo has formed; it is possible that the particles in that shell pass through the origin and then continue drifting out of the overdense region.  To determine the outcome of gravitational collapse during radiation domination, we used the cosmological simulation code \texttt{GADGET-2} \cite{2005MNRAS.364.1105S} to simulate the evolution of overdense regions.  We modified the code to include both radiation \cite{DEAB18a, DEAB18b} and decaying $Z'$ particles.  These components are modeled at the homogeneous level by adding corresponding terms to the Hubble rate appearing in the time integrals used for particle drifts and gravitational kicks.  The clustering capacity of the $Z'$ particles is modeled by scaling the mass, $m$, of each simulation particle as 
\begin{equation}\label{mass}
m = \left(\frac{\bar{\rho}_{Z'} + \bar{\rho}_X}{\bar{\rho}_X}\right) m_0,
\end{equation}
where $m_0$ is the particle mass at late times \cite{1987ApJ...321...36S,Enqvist:2015ara,Dakin:2019dxu}.  Effectively, each simulation particle represents $Z'$ and $X$ particles together.\footnote{As a technical matter, this modification is also implemented by modifying a time integral. In particular, the time integrand used for gravitational kicks is scaled by $m/m_0$.}  We verified that this simulation matches the predictions of linear theory while $\delta \ll 1$.  

The volume of our simulation box is  ($10^{-9}h^{-1}$ Mpc)$^3$ and we start the simulation at $a=3\times 10^{-13}$ in a scenario with $T_\mathrm{RH}=22$ MeV, which implies that the EMDE ends and radiation domination begins at $a = 10^{-11}$.
We assume that the dimensionless power spectrum scales as $\mathcal P(k) \propto \ln(k/k_0)^2$ with \mbox{$k_0\simeq 6\times 10^6 h\,\mathrm{Mpc^{-1}}$} for the relevant perturbation wavenumbers; that is, we consider scales that entered the horizon during a period of radiation domination prior to the EMDE.  We then draw individual proto-halo density peaks from the Gaussian-filtered power spectrum,
\begin{equation}
\mathcal{P}_f(k)=\mathcal{P}(k)\exp(-R_f^2 k^2),
\label{filtered}
\end{equation}
using the density profiles of Ref.~\cite{BBKS86} with smoothing scale $R_f=10^{-8}h^{-1}$ kpc.  To add some realism to this otherwise smooth density peak, we also superpose small-scale Gaussian noise drawn from the oppositely filtered power spectrum.  

\begin{figure}[t]
\resizebox{3.4in}{!}
{
      \includegraphics{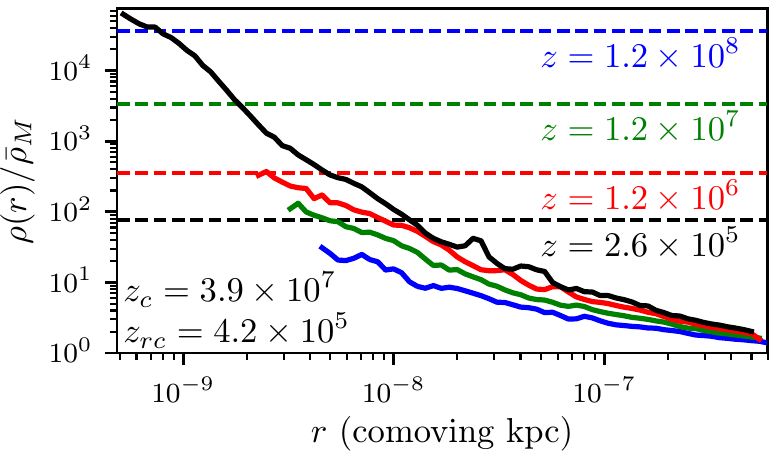}
 }
\caption{The evolution of an overdense region that collapses during radiation domination.  The solid lines show the spherically averaged matter density profile at four different redshifts.  The minimum radius of the profile at each redshift corresponds to a sphere that contains 100 particles.  The dotted lines show the density of the radiation at these same redshifts.  The spherical-collapse model predicts that this overdense region should collapse at $z_c=3.9\times10^7$, but the matter density grows slowly until the matter density within the smoothing scale ($R_f=10^{-8}h^{-1}$ kpc) exceeds the radiation density, which occurs at $z_{rc} = 4.2 \times 10^5$.  At that point, the central density increases dramatically, forming a bound object with a central density that exceeds the background matter density by over four orders of magnitude.}
\label{Fig:recollapse}
\end{figure}

We simulated the evolution of overdense regions with several different initial density profiles, three-dimensional shapes, and collapse redshifts, $z_c$, where $1+z_c = a_c^{-1}$ as predicted by the spherical collapse model.  For overdense regions that collapse well after reheating, these simulations revealed a universal evolution that varied very little between the simulations.  First, the density profile of the overdense region developes a shallow cusp with ${\mathrm d \ln \rho / \mathrm d \ln r \simeq -1}$ at $z\simeq z_c$.  The density profile of this proto-halo is well-fit by the function
\begin{equation}
\rho(r) = \left[ A \ln\left(\frac{1+z_c}{1+z}\right) + B\right] \bar{\rho}_M \, \frac{R_f}{r},
\label{protohalorho}
\end{equation}
where $A \simeq 8.4$, $B \simeq 12.4$, and $\bar{\rho}_M$ is the cosmological background matter density.
When the average matter density within $R_f$ exceeds the homogeneous radiation density, the proto-halo re-collapses to form a much denser halo.  This evolution is depicted in Fig.~\ref{Fig:recollapse}.   In contrast, overdense regions that collapse too soon after reheating do not re-collapse to form a halo because the dark matter particles are moving fast enough to pass through the overdense region and escape the gravitational forces induced by the growing matter density.  For $\Trh \gsim 10$ MeV, our simulations indicate that regions with $z_c \lsim 10^{8}$ re-collapse.

Since our simulations only include a single overdense region with vacuum boundary conditions, our simulated halos do not develop the double power-law density profiles seen in cosmological simulations that start from a Gaussian density field.  However, since the re-collapse of the proto-halo occurs in a locally matter-dominated environment, is seems reasonable to assume that the resulting halo should be similar to a halo that forms shortly after matter-radiation equality.  Refs.~\cite{DEAB18a, DEAB18b} showed that halos forming at $z\simeq 1000$ in cosmological simulations with Gaussian initial conditions realized from an enhanced scale-invariant power spectrum have Navarro-Frenk-White (NFW) profiles \cite{NFW97}.  We assume that the halos that form during radiation domination also develop NFW profiles and that the concentration parameter of the resulting halo is determined by the redshift of re-collapse, not $z_c$.  We use the density profile given in Eq.~(\ref{protohalorho}) to determine the re-collapse redshift, $z_{rc}$, from $z_c$ ($z_{rc}$ is the redshift at which the average matter density within $R_f$ exceeds the radiation density, unless that is already true at $z_c$, in which case $z_{rc}=z_c$). To be conservative, we only consider $z_c \lsim 10^6$, thereby ensuring that the collapsing region forms a gravitationally bound halo (as demonstrated in Fig.~\ref{Fig:recollapse} for  $z_c=3.9\times10^7$).  Restricting ourselves to halo formation with $z_c <10^6$ also gives the halos extra time to establish NFW profiles after they re-collapse, and we explore how possible further delays in halo formation affect our gamma-ray constraints in Sec.~\ref{results}.

\section{Structure Formation during the EMDE: The Impact of Gravitational Heating}
\label{heating}

If the duration of the EMDE is sufficiently long, then the density perturbations among the $X$ and $Z'$ particles will become large enough to form gravitationally bound structures prior to reheating.  The characteristic initial density perturbation is $\delta \simeq 10^{-5}$ at horizon entry.  Both $\delta_X$ and $\delta_{Z'}$ grow by roughly a factor of 100 at horizon entry and then linearly with the scale factor during the EMDE (as seen in Fig.~\ref{Fig:growth}), so we expect significant structure formation if $a_\mathrm{RH}/a_\mathrm{dom} \gsim 10^{3}$.  
Since $\rho_{Z'} \propto a^{-3}$ during the EMDE and $\rho_{Z'}(a_\mathrm{dom})=\rho_\mathrm{SM}(a_\mathrm{dom})$ and $\rho_{Z'}(a_\mathrm{RH}) \simeq \rho_\mathrm{SM}(a_\mathrm{RH})$, $a_\mathrm{RH}/a_\mathrm{dom} \simeq(g_{*,\mathrm{dom}}/g_{*,\mathrm{RH}})^{1/3} (T_\mathrm{dom}/\Trh)^{4/3}$.  Therefore, we expect that most of the hidden sector particles will be in halos at reheating if \mbox{$T_\mathrm{dom} \gsim 200 \, \Trh$}.  

A more precise determination of the halo abundance can be obtained from the Press-Schechter formalism \cite{PS74}, which provides an expression for the the fraction of dark matter that is bound into halos as a function of halo mass:
\beq
\frac{d f}{d \ln M} = \sqrt{\frac{2}{\pi}} \left|{\frac{d \ln\sigma}{d \ln M}}\right| \frac{\delcoll}{\sigma(M,z)} \exp\left[-\frac{\delcoll^2}{2\sigma^2(M,z)}\right],
\label{dfdlnM}
\eeq
where $\sigma^2(M,z)$ is the variance of $\delta \rho/\bar{\rho}$ in spheres that contain mass $M$ in $X$ and SM particles evaluated using a sharp-$k$ filter at \mbox{$k_s = 2.5/[3M/(4\pi \rho_{m,0})]^{1/3}$}, as advocated by Ref. \cite{2013MNRAS.428.1774B}.  Since we are interested in structure formation during a matter-dominated era, $\delta_c = 1.686$.  We integrate Eq.~(\ref{dfdlnM}) over all halo masses to obtain the total fraction of matter contained in halos at reheating.  This computation confirms that halos are common for  \mbox{$T_\mathrm{dom} \gsim 200 \, \Trh$}. For example, if $T_\mathrm{dom} = 470$ GeV and $\Trh = 1$ GeV, 20\% of the matter is bound into halos at the time of reheating.  

The halos present at reheating are predominantly composed of $Z'$ particles; while $H \gg \Gamma_{Z'}$, ${\rho_X}/{\rho_{Z'}} \simeq T_\mathrm{eq}/T_\mathrm{RH}$.  When the $Z'$ population decays into relativistic SM particles, their energy will leave the halo.  The remaining $X$ particles will escape the halo if their orbits cannot adiabatically adjust to the reduction in the halo's mass.  The characteristic dynamical time of a halo is $t_\mathrm{dyn} \simeq (G\rho)^{-1/2}$, while the characteristic mass-loss timescale is $t_{\Delta M} = M/\dot{M} = (\rho_X + \rho_{Z'})/(\Gamma_{Z'} \rho_{Z'})$.  For a halo that forms at $a_f$, the ratio of these timescales is as follows: 
\begin{align}
\frac{t_{\Delta M}}{t_\mathrm{dyn}} \simeq& \frac{\sqrt{G\rho_{Z'}}}{\Gamma_{Z'}} \left(\frac{\rho_X}{\rho_{Z'}}e^{\Gamma_{Z'}\Delta t} + 1\right) \!\!\sqrt{\frac{\rho_X}{\rho_{Z'}}+e^{-\Gamma_{Z'}\Delta t}} \\
\simeq& \sqrt{\frac{3}{8\pi}}\left[\frac{a_\mathrm{RH}}{a_f}\right]^{3/2}\!\!\left(\frac{T_\mathrm{eq}}{\Trh}e^{\Gamma_{Z'}\Delta t} + 1\right) \nonumber\\
&\times\sqrt{\frac{T_\mathrm{eq}}{\Trh}+e^{-\Gamma_{Z'}\Delta t}}, \nonumber
\end{align}
where $\Delta t$ is the time since the halo's formation. This ratio falls below unity for all halos with 
\beq
\frac{a_f}{a_\mathrm{RH}} \gsim \left(\frac{81}{32\pi} \frac{T_\mathrm{eq}}{\Trh} \right)^{1/3}.
\eeq
Therefore, we expect that all halos that form at scale factors greater than $10^{-6} a_\mathrm{RH}$ will be destroyed at reheating and that the $X$ particles within these halos will be released to free stream with the velocity they had within the halo.  In this way, the formation of structure during the EMDE effectively heats the dark matter, imparting the $X$ particles with much larger velocities than they otherwise would have had. 

The typical particle velocity within a halo is given by the virial velocity dispersion, $\sigma_v \simeq \sqrt{GM/R}$, where $R$ is radius within which the average density is 200 times the background density and $M$ is the mass enclosed in this radius.  From this, it follows that
\begin{align}
\sigma_v &\simeq \sqrt{G\left(\left.\frac{\rho_{Z'}}{\rho_X}\right|_{a_f}M_X\right)^{2/3}\left[\frac{4\pi}{3}200 \rho_{X+Z'}(a_f)\right]^{1/3}} \nonumber \\ \label{sigv}
&\propto \sqrt{\frac{M_X^{2/3}}{a_f}},
\end{align}
where $M_X$ is the mass in $X$ particles.  The formation time of a halo is determined by the amplitude of density fluctuations with $k \simeq [3M_X/(4\pi \rho_{X,0})]^{-1/3}$; halos with this mass will typically form when $\delta \simeq \delta_c$ for this wavenumber.  For modes that enter the horizon during an EMDE, $\delta_X \propto k^2$ \cite{ES11}.  Since these modes grow linearly with the scale factor during the EMDE, $a_f \propto k^{-2} \propto M_X^{2/3}$ and $\sigma_v \simeq 2000$ km/s, independent of halo mass.  The amplitudes of modes that enter the horizon before the EMDE starts do not increase as rapidly with increasing $k$ because these modes only grow logarithmically between horizon entry and the start of the EMDE.  As a result, $a_f$ does not decrease as fast with increasing $M_X$, and $\sigma_v$ increases as $M_X$ increases.  Therefore, the largest halos that form during the EMDE will always have the largest virial velocities.  

After the dark matter particles are released from the halos that evaporate at reheating, they will free stream in random directions, thereby suppressing any density perturbations with wavelengths less than the free streaming horizon, $\lambda_\mathrm{fs}$.  If all the dark matter particles have the same velocity dispersion, the matter power spectrum is suppressed by a factor of $\exp[-k^2\lambda_\mathrm{fs}^2]$ \cite{GHS05, Bertschinger06, BLRV09}, where 
\beq
\lambda_\mathrm{fs} = \int^{t_0}_{t_\mathrm{kd}} \frac{v}{a} \drm t = \sigma_v a_\mathrm{RH} \int_{a_\mathrm{RH}}^{1} \frac{\drm a}{a^3 H(a)}.
\label{lambdafs}
\eeq
To probe how this gravitational heating of dark matter affects the formation of microhalos after the EMDE, we make two estimates.  For the ``optimistic" estimate, we only consider the coldest 80\% of the dark matter.  If less than 20\% of the $X$ particles are bound into halos at reheating, we impose no free streaming cutoff to the matter power spectrum.  Otherwise, we find the halo mass, $M_{20}$, such that 20\% of the $X$ particles are bound into halos with $M>M_{20}$ at reheating.  At $a=0.8 \, a_\mathrm{RH}$ about 10\% of the matter is bound into such halos, so we take this to be the formation time of these halos.  We then use Eq.~(\ref{sigv}) to determine the virial velocity within a halo with a mass equal to $M_{20}$ that formed at $a_f=0.8 \, a_\mathrm{RH}$ and apply a $\exp[-k^2\lambda_\mathrm{fs}^2]$ cutoff to the matter power spectrum with $\lambda_\mathrm{fs}$ given by Eq.~(\ref{lambdafs}).  We also assume that only 80\% of the dark matter participates in structure formation after the EMDE and reduce the growth rate of perturbations accordingly.  For the ``pessimistic" estimate, we follow this same procedure for the coldest 90\% of the dark matter. Since $M_{10}>M_{20}$, this procedure yields a larger value of $\sigma_v$ and suppresses perturbations on larger scales.  

Both of these estimates are conservative in that most of the dark matter does not free stream to this extent, but it is also possible that the actual cut-off in the matter power spectrum differs from $\lambda_\mathrm{fs}$.  Unfortunately, the boost factor from microhalos that form after an EMDE is extremely sensitive to the cut-off scale in the matter power spectrum; changing this scale by a factor of two can change the boost factor by more than an order of magnitude \cite{E15}.  The contrast between the optimistic and pessimistic estimates gives some idea of the uncertainty associated with gravitational heating, but it is certainly possible that both approaches overestimate how much gravitational heating suppresses the microhalo population.  More comprehensive numerical simulations of halo evolution during and after the EMDE could allow us to extend our constraints to larger values of $T_\mathrm{dom}/T_\mathrm{RH}$.

\section{Microhalo-dominated Dark Matter Annihilation}
\label{results}


If most of the dark matter is concentrated within a population of microhalos, then the dark matter annihilation rate in a given volume of space will be given by
\beq
\Gamma = \frac{\sigv}{2 m_X^2}  \sum_i \int \rho_{X,i}^2(\vec{r}) \,\drm^3 \vec{r}, 
\label{annrate}
\eeq
where the sum is over the microhalos within the volume and $\rho_{X,i}(\vec{r})$ is the density profile of microhalo, $i$. If, for illustration, we make the simplifying assumption that each microhalo is identical, we find that the dark matter annihilation rate is proportional to the number of microhalos, $N_h$, within that volume:
\beq
\Gamma =  \frac{\sigv}{2 \mdm^2} N_h J_h,
\eeq
where $J_h \equiv \int \rho_{X,h}^2(\vec{r}) \,\drm^3 \vec{r}$. 

To determine the fraction of dark matter that is bound into halos as a function of halo mass, we use the Press-Schechter formalism \cite{PS74}, as described in the previous section.
If the halos present at a redshift $z_f$ survive to the present day, then the annihilation rate in a volume will be given by
\beq
\Gamma = \frac{\sigv}{2 m_X^2} \int \drm^3 \vec{r} \int_{0}^{\mdec}  \drm \ln M_f \,\frac{\rho_m(\vec{r})}{M_f} \,\left.\frac{d f}{d \ln M}\right|_{z_f} J_h(M_f),
\label{annratemicropop}
\eeq
where $\rho_m$ is the matter density and $M_{\rm RH}$ is the mass enclosed in the cosmological horizon at reheating.  We only consider the microhalos with $M<\mdec$ because these microhalos form much earlier than larger microhalos and are therefore more likely to survive accretion into larger halos \cite{E15}.

If the microhalos present at redshift $z_f$ have NFW profiles with concentration $c = r_\mathrm{200}/r_s$, where $r_\mathrm{200}$ is the radius within which the average matter density is 200$\bar{\rho}_M$, then
\begin{eqnarray}
J_h &=&\frac{f_X^2M_f}{3}200\bar{\rho}_{M}(z_f)\left[\frac{c^3}3-\frac{c^3}{3(1+c)^3}\right] \nonumber \\ 
&\times& \left[\ln(1+c)-\frac{c}{1+c}\right]^{-2}, 
\label{J_h}
\end{eqnarray}
where $f_X$ is the fraction of mass that is dark matter.  We assume that all microhalos have $c=2$ at a redshift $z_f$ because this is the lowest concentration seen in microhalo simulations \cite{AD12, Ishiyama14}.  We will revisit this assumption when we discuss uncertainties in our computation of gamma-ray emission later in this section.

Inserting Eq.~(\ref{J_h}) into Eq.~(\ref{annratemicropop}) gives
\begin{align}
\Gamma =& \frac{\sigv}{2 m_{X}^2}  \left(\int_V f_X \rho_m(\vec{r}) \drm^3 \vec{r} \right) \frac{200}{3}f_X \bar{\rho}_{M,0}(1+z_f)^3  \nonumber \\
&\times13.76\int_{0}^{\mdec} \left.\frac{d f}{d \ln M}\right|_{z_f} \drm \ln M_f. \
\end{align}
If we define $f_\mathrm{tot}$ to be the fraction of mass contained in halos with $M<\mdec$, the annihilation rate in a volume divided by the dark matter mass within that volume is
\beq
\frac{\Gamma}{M_X} = \frac12 \left(\frac{\sigv/m_X^2}{\mathrm{GeV}^{-4}}\right) B_0 \times (8.098\times10^{-47}) \, \Omega_X h^2,
\label{annratepermass}
\eeq
where $B_0 = 13.76\times{200}(1+z_f)^3 f_\mathrm{tot}(z_f)/3$ is the boost due to the microhalos.  As in Ref.~\cite{E15}, we choose $z_f$ such that $B_0$ is maximized.  While this estimate of the boost factor assumes that all the microhalos present at a redshift of $z_f$ survive until the present day, it also neglects the annihilation boost due to any subhalos present within these microhalos and any microhalos that form at $z<z_f$.  Since $B_0$ only depends on the fraction of dark matter bound into microhalos, and not on the size of the microhalos, it does not depend strongly on the reheat temperature, but it is very sensitive to the ratio $\kcut/\krh$, where $\kcut$ is the scale that sets the cut-off in the matter power spectrum, $P(k) \propto \exp[-k^2/\kcut^2]$ ~\cite{E15}.  In the absence of gravitational heating, $\kcut$ is the scale that enters the horizon when $T_h = m_{Z'}$. Otherwise it is the free-streaming scale determined by the velocities of the $X$ particles.

\begin{figure}[t]
\includegraphics[scale=0.47]{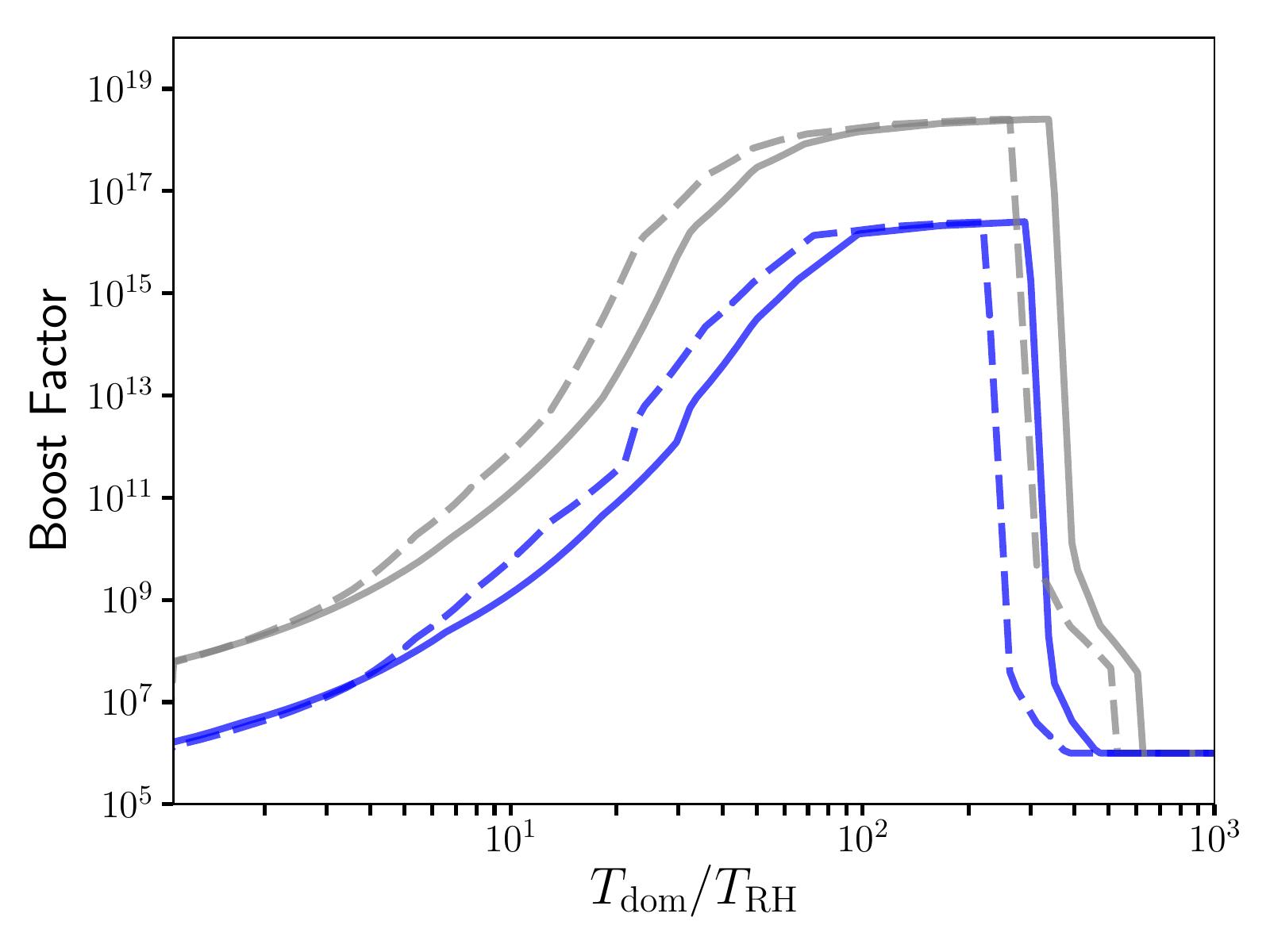} 
\caption{The annihilation boost factor, $B_0$, as a function of the ratio of the temperatures at the beginning of the EMDE, $T_{\rm dom}$, and at the time of reheating, $T_{\rm RH}$, for $\alpha_X=0.01$ (dashed) and $\alpha_X=0.1$ (solid). For each curve, we adopt $\xi_{\rm inf}=1$, $m_X/m_{Z'}=20$ and select $\epsilon$ and $m_X$ such that $\Omega_X h^2 \simeq 0.11$. The grey and blue curves represent the results that are found following our optimistic and conservative procedures, respectively. In scenarios with a very long EMDE, gravitational heating significantly disrupts the microhalo population, reducing the boost factors to values near that predicted for an ordinary thermal relic, $B_0 \sim 10^6$.}
\label{boostratio}
\end{figure}

In Fig.~\ref{boostratio}, we plot the value of the boost factor, $B_0$, as a function of $T_{\rm dom}/T_{\rm RH}$, for the case of $\alpha_X=0.01$ (dashed) and $\alpha_X=0.1$ (solid). At each point, we adopt $\xi_{\rm inf}=1$, $m_X/m_{Z'}=20$, and select the values of $\epsilon$ and $m_X$ such that we obtain the desired dark matter abundance, $\Omega_X h^2 \simeq 0.11$. In each case the grey and blue curves represent the results found following our default and conservative procedures, respectively (as described later in this section).  Applying our computation of $B_0$ to a power spectrum derived assuming no deviations from radiation domination in the early universe gives $B_0 \simeq 10^6$ for a very broad range of minimum halo masses.  In scenarios with a very long EMDE, gravitational heating significantly suppresses the microhalo population, reducing the boost factors to this value.  We also see in Fig.~\ref{boostratio} that very short EMDEs still increase the boost factor: the presence of the $Z'$ particles enhances the growth of dark matter perturbations even if they never fully dominate the energy content of the universe. 

The difference between the solid ($\alpha_X=0.01$) and dashed ($\alpha_X=0.1$) curves in Fig.~\ref{boostratio} reflects the fact that $B_0$ depends on both $T_\mathrm{RH}$ and $T_\mathrm{dom}$.  The enhancement to small-scale perturbations due to the EMDE only depends on the ratio $k_\mathrm{dom}/k_\mathrm{RH}$, where $k_\mathrm{dom}$ is the wave number of the mode that enters the horizon when $\rho_{Z'} = \rho_R$ and $k_\mathrm{RH}$ is the wavenumber of the mode that enters the horizon at reheating, as defined in Ref. \cite{E15}.  This ratio is determined by $T_\mathrm{dom}$ and $T_\mathrm{RH}$:
\beq
\frac{k_\mathrm{dom}}{k_\mathrm{RH}} = \sqrt{2} \left[\frac{g_\star(T_\mathrm{dom})}{g_\star(T_\mathrm{RH})}\right]^{1/6} \left(\frac{T_\mathrm{dom}}{T_\mathrm{RH}}\right)^{2/3},
\eeq
for ${T_\mathrm{dom}}/{T_\mathrm{RH}} \gtrsim 5$. For fixed $k_\mathrm{dom}/k_\mathrm{RH}$, $B_0$ still has a weak dependence on $T_\mathrm{RH}$ due to the red tilt of the primordial power spectrum and the duration of the radiation-dominated era, but this slight variation is less than a factor of 5 for the range of $T_\mathrm{RH}$ we consider.

Since the microhalos are expected to track the dark matter density (as opposed to the square of the density), the emission signatures in these scenarios are more akin to decaying dark matter than traditional annihilating dark matter.  With this in mind, we will use measurements of the high-latitude gamma-ray background to constrain these models.  The annihilation rate per mass given by Eq.~(\ref{annratepermass}) can be converted to an effective lifetime by considering the rate of jet production in a fixed volume containing dark matter mass, $M_X$.\footnote{We use the phrase ``jet production'' here to denote the production of any primary annihilation product, and can include leptons which do not result in the formation of a QCD jet.}  Dark matter annihilations produce jets of energy $m_X$ at a rate of $2(\Gamma/M_X)M_X$.  To get the same jets from decaying dark matter, the particles must have mass $2 m_X$, which implies that the jet production rate is $(2/\tau)[M_X/(2 m_X)]$.  Equating these rates yields the following effective lifetime:
\beq
\tau_\mathrm{eff} = \frac{1}{2m_X(\Gamma/M_X)}.
\eeq
This lifetime should be compared to the lower bound on the dark matter lifetime for particles with mass $2 m_X$.

In Fig.~\ref{limits}, we plot the effective lifetime of the dark matter as a function of $m_X$, for the same parameter values (and the same line types) as adopted in Fig.~\ref{boostratio}. These curves do not extend to arbitrarily low $m_X$ values because we only show results for $k_{\rm dom}/k_{\rm RH} > 1.5$, which corresponds to requiring that the density of $Z'$ particles exceeds the density of SM particles at some point prior to their decay.  We limit our analysis to scenarios with an EMDE because the annihilation rate scales with the number density of microhalos (and thus with the density of dark matter) only if the boost factors are very large.  In a scenario without an EMDE, constraints based on gamma-ray observations of dwarf spheroidal galaxies~\cite{Fermi-LAT:2016uux} should instead be applied to the model.

\begin{figure}[t]
\includegraphics[scale=0.47]{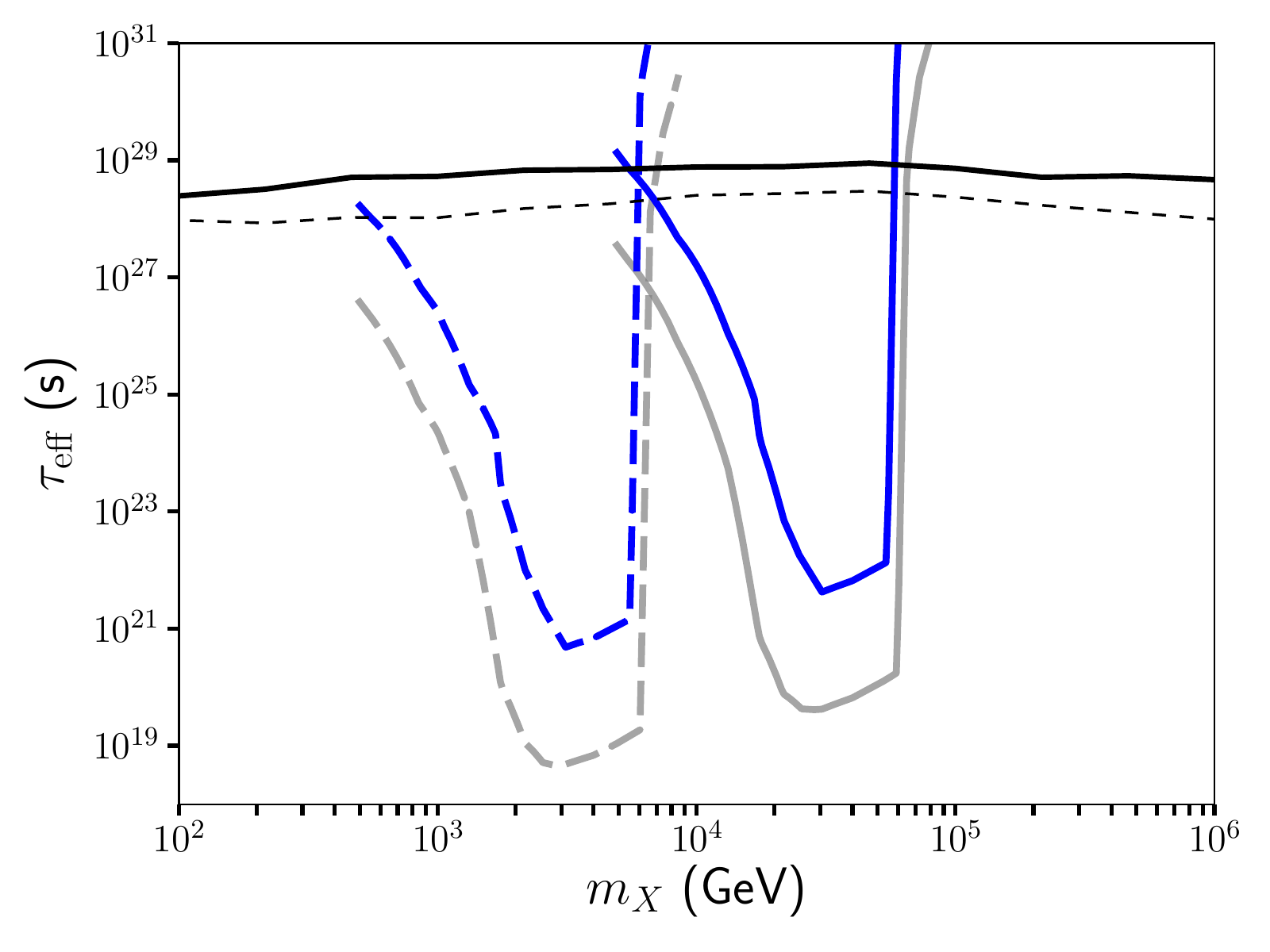} 
\caption{The dark matter's effective lifetime as a function of mass, for the same parameters as adopted in Fig.~\ref{boostratio}. Again, the grey and blue curves represent the results found following our optimistic and conservative procedures, respectively. Results are shown only for $T_\mathrm{dom}/T_\mathrm{RH} \gtrsim 1.1$, and thus these curves do not extend to arbitrarily low masses. The black curves represent the lower limits on $\tau_{\rm eff}$ in this model, as derived from Fermi's measurement of the high-latitude gamma-ray background~\cite{Ackermann:2014usa,Blanco:2018esa}. These constraints rule out scenarios with an EMDE for dark matter masses lighter than $m_X \lsim 6-7$ TeV for $\alpha_X=0.01$ and lighter than $\sim 60-70$ TeV for $\alpha_X=0.1$. For higher masses, gravitational heating significantly disrupts the microhalo population, reducing the boost factors to acceptable levels.}
\label{limits}
\end{figure}  

The uncertainties in our computation of the boost factor, $B_0$, arise from two main sources: the cut-off in the matter power spectrum ($P(k) \propto \exp[-k^2/k_\mathrm{cut}^2]$) and the characterization of the halo density profiles.  The first cut-off scale we consider is set by the mass of the $Z'$ particle: $k_\mathrm{cut} = k_{Z'}$ where $k_{Z'}$ is the wavenumber of the mode that enters the horizon when the temperature of the hidden sector is equal to $m_{Z'}$.  This cut-off only approximates the suppression of perturbations on scales that enter the horizon while the $Z'$ population is relativistic and it is possible that the actual power spectrum has a slightly different shape.  This change will not affect scenarios with $30\lsim k_\mathrm{dom}/k_\mathrm{RH} \lsim 70$ because most of the dark matter is already bound into microhalos at $a=10^{-6}$ in these scenarios, so changing the power spectrum has a minimal effect on the value of $B_0$ obtained from the halo population at this time.  

If modes that enter the horizon while $T_h\simeq m_{Z'}$ are not as damped as this model predicts, $B_0$ will increase for  $k_\mathrm{dom}/k_\mathrm{RH} \lsim 30$ because microhalos will form earlier.  However, more structure will also form prior to reheating, so gravitational heating will suppress $B_0$ at smaller values of $k_\mathrm{dom}/k_\mathrm{RH}$.  For example, if $k_\mathrm{cut} = 2k_{Z'}$, gravitational heating will suppress $B_0$ for $k_\mathrm{dom}/k_\mathrm{RH} \gsim 70$ instead of $k_\mathrm{dom}/k_\mathrm{RH} \gsim 80$.   Conversely, if $k_\mathrm{cut} = k_{Z'}/2$, the impact of gravitational heating will be pushed to larger values of  $k_\mathrm{dom}/k_\mathrm{RH}$, but $B_0$ will be suppressed by about an order of magnitude for $k_\mathrm{dom}/k_\mathrm{RH}\lsim 30$ due to the later formation of microhalos.  For the conservative $B_0$ calculation, we adopt $k_\mathrm{cut} = k_{Z'}/2$ for $k_\mathrm{dom}/k_\mathrm{RH}\lsim 30$ to account for this suppression.

The second cut-off scale we consider arises from the suppression of perturbations following gravitational heating.  We computed this suppression based on the evolution of the coldest 80\% and 90\% of the dark matter to illustrate the uncertainty associated with structure formation durning the EMDE, but it is possible that both of these calculations overestimate the suppression of perturbations due to gravitational heating.  Simulations of halo evaporation at reheating could reveal that the free-streaming of particles released from halos does not result in $k_\mathrm{cut} = \lambda_{\mathrm{fs}}^{-1}$ as we have assumed.  For scenarios with gravitational heating, $k_\mathrm{RH}\lambda_{\mathrm{fs}}^{-1} \simeq 40$; changing $k_\mathrm{cut}$ by a factor of two can change $B_0$ by up to two orders of magnitude but would still not alter our constraints.  Determining the onset of gravitational heating is a more significant source of uncertainty.  In our optimistic estimate, gravitational heating dramatically suppresses $B_0$ for $k_\mathrm{dom}/k_\mathrm{RH} \gsim 90$ because $20\%$ of the dark matter is gravitationally heated in these scenarios. It is worth noting, however, that less than half of the dark matter is gravitational heated for $k_\mathrm{dom}/k_\mathrm{RH} \lsim 130$.  Therefore, it is possible that the formation of microhalos in these scenarios is not completely suppressed by the free streaming of released dark matter particles.  For our default $B_0$ computation, we base the free streaming horizon on the coldest $80\%$ of the dark matter, while the conservative computation uses the coldest 90\% of the dark matter.  The difference in $B_0$ between these two approaches is roughly equivalent to increasing $k_\mathrm{cut}$ to $2k_{Z'}$ while basing the free-streaming horizon on the coldest 80\% of the dark matter.

Turning to the characterization of the halo density profiles, our procedure for calculating $B_0$ relies on the assumption that the boost factor is dominated by halos present at a certain redshift and that all halos have NFW profiles with concentration $c=2$ at this time.  We choose the redshift that maximizes $B_0$ by striking the optimal compromise between increased microhalo abundance and increased microhalo density.  We then assume that these halos survive to the present day and that any halos that form subsequently make a minimal contribution to the boost factor.  Remarkably, the boost factor computed by this simple approach applied to a standard matter power spectrum (i.e. no deviations from radiation domination at early times) nearly matches the results of more complicated halo-model approaches that use concentration-mass relations and (sub)halo mass functions calibrated using $N$-body simulations. The $\zeta(z)$ boost factor used in Ref. \cite{2015JCAP...09..008T} is related to $B_0$ by $B_0 = \zeta(z)(1+z)^3$.  For a minimum halo mass of $10^{-6} h^{-1}\,M_\odot$, $\zeta(z)(1+z)^3 \simeq 4\times10^5$, whereas our method applied to a standard power spectrum yields $B_0 \simeq 10^{6}$.   The discrepancy of a factor of $\sim2-3$ could easily be attributed to the destruction of subhalos.

Nevertheless, we should question the density profiles assumed by our calculation.  The adoption of an NFW profile is probably overly conservative, as numerous simulations have indicated that the first generation of halos have steeper inner density profiles \cite{AD12, Ishiyama14, DEAB18a, DEAB18b, DBA19}.  However, simulations also show that it takes considerable time after the halo's formation for the NFW profile to stabilize: a halo that forms at $z=1000$ from initial conditions drawn from an enhanced plateau in the power spectrum has an NFW profile with $c=2$ at $z\simeq400$ \cite{DEAB18a, DEAB18b}.  The time interval between these two redshifts corresponds to $10 \, t_\mathrm{dyn}$, where $t_\mathrm{dyn} = \sqrt{3\pi^2/(16G\rho_\mathrm{vir})}$ with $\rho_\mathrm{vir} =200 \bar{\rho}_M(z_f)$ is the dynamical time of the halo at formation.  Waiting ten dynamical times after $z_f$ to evaluate the density profile reduces $B_0$ by a factor of 10 for halos that form during matter domination, which would apply to $k_\mathrm{dom}/k_\mathrm{RH} \lsim 10$.  For halos that form during radiation domination, the reduction factor depends on $z_f= z_\mathrm{rc}$, increasing to 100 for $a_c = 10^{-6}$.  However, if most of the dark matter is bound into halos at $a = 10^{-6}$, as the case for $30\lsim k_\mathrm{dom}/k_\mathrm{RH} \lsim 70$, then it is safe to assume that halos were present at much earlier times and that our estimate of $B_0$ based on the halos present at $a_f = 10^{-6}$ underestimates the true annihilation rate.  Nevertheless, for our conservative $B_0$ computation, we reduce all $B_0$ values by the reduction factor implied by delaying the establishment of the NFW profile with $c=2$ ten dynamical times after the halos form.

The most stringent and broadly applicable constraints on this class of models are derived from the measurements of the Fermi Gamma-Ray Space Telescope. To derive these constraints, we follow the procedure described in Ref.~\cite{Blanco:2018esa}, modified as described in Appendix~\ref{gamma} of this paper. The black curves shown in Fig.~\ref{limits} represent the lower limits on $\tau_{\rm eff}$ in this model, as derived from Fermi's measurement of the high-latitude gamma-ray background~\cite{Ackermann:2014usa}. The dashed (solid) curves derive from the default (more conservative) procedure as described in Ref.~\cite{Blanco:2018esa}. When these constraints are compared to the predicted values of $\tau_{\rm eff}$, we find that a wide range of dark matter masses are already ruled out. In particular, for the case of $\alpha_X=0.01$ (0.1), this excludes any scenario with a significant EMDE for masses up to $m_X \sim 6-7$ TeV (60-70 TeV). For higher masses, gravitational heating suppresses the abundance of microhalos, reducing the boost factors to a level consistent with Fermi's measurements.

\begin{figure}[t]
\includegraphics[scale=0.47]{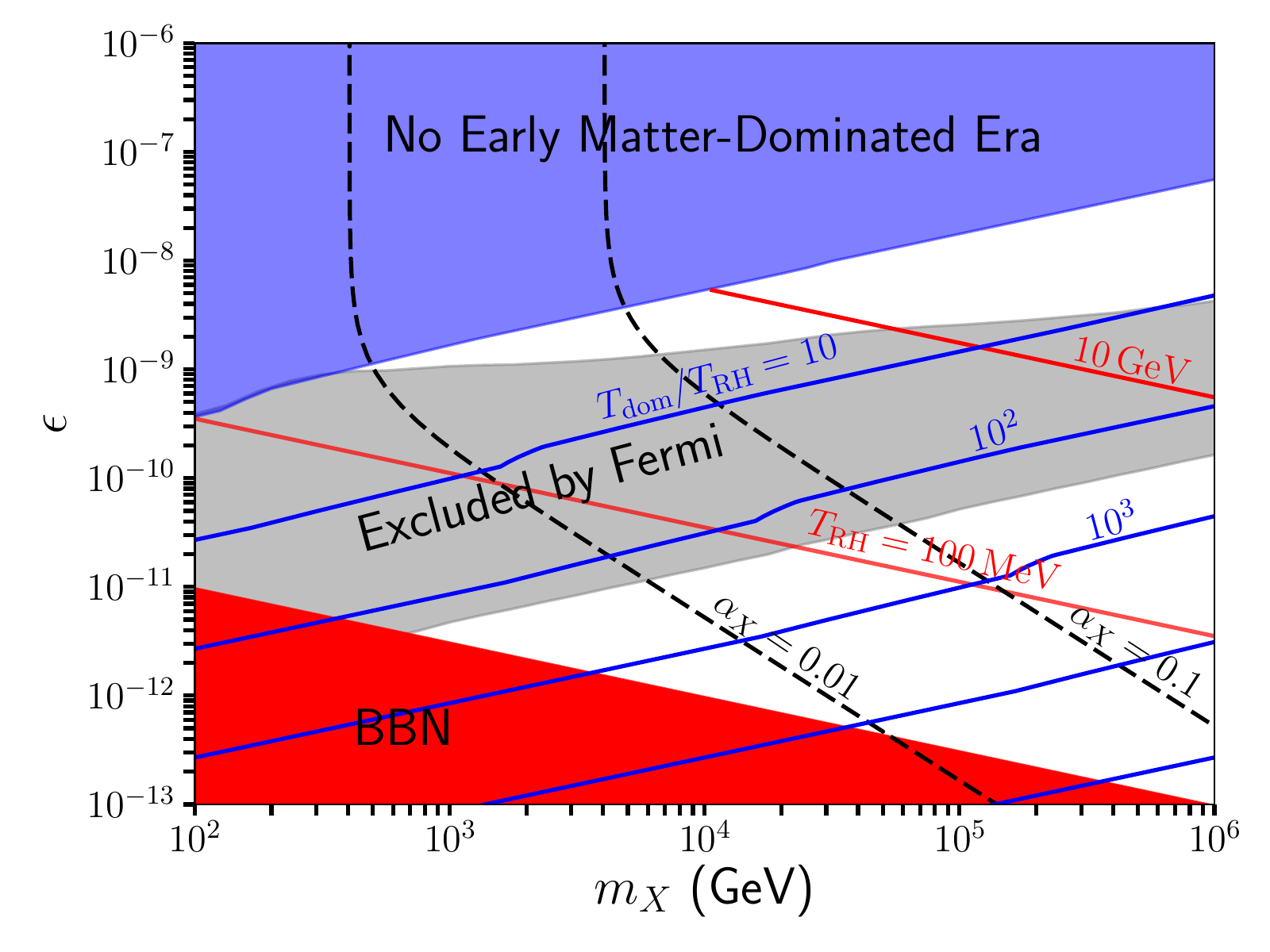} 
\caption{Some of the features and constraints found across the parameter space of the vector portal dark matter model, for the case of $\xi_{\rm inf}=1$ and $m_X/m_{Z'}=20$. The dashed curves represent the regions of this plane which yield a relic abundance equal to the measured dark matter density, $\Omega_X h^2 \simeq 0.11$, for choices of $\alpha_X=0.01$ and $\alpha_X=0.1$. In the blue region, the hidden sector never dominates the energy density of the early universe, and the red region is excluded by the measured light element abundances and other cosmological considerations. 
Throughout the grey region, the EMDE leads to the formation and survival of a large population of microhalos, resulting in large boost factors (following our conservative procedure) that are currently ruled out by Fermi's measurement of the high-latitude gamma-ray background. In the lower right portions of this figure, gravitational heating leads to the suppression of this microhalo population, reducing the boost factors and resulting gamma-ray emission to acceptable levels.  Note that in the lowest portions of the region labeled ``No Early Matter-Dominated Era'', the energy density of the $Z'$ population is still sizable, leading to non-negligible contributions to the annihilation boost factor.}
\label{plane}
\end{figure}  

In Fig.~\ref{plane}, we plot some of the features of and constraints on this model, again for the case of $\xi_{\rm inf}=1$ and $m_X/m_{Z'}=20$. The dashed curves represent the regions of this plane that yield a relic abundance equal to the measured dark matter density, $\Omega_X h^2 \simeq 0.11$, for choices of $\alpha_X=0.01$ or $\alpha_X=0.1$. In the upper regions of this plane, the hidden sector never dominates the energy density of the early universe and there is no EMDE. For smaller values of $\epsilon$, however, the $Z'$ population is sufficiently long-lived that it comes to dominate the energy density and subsequently reheat the SM bath through its decays. The blue curves represent contours of constant $T_{\rm dom}/T_{\rm RH}$ (see Eq.~\ref{TdomTRH}), while the red curves are constant in $T_{\rm RH}$ (see Eq.~\ref{TRH}). The lower left region of the plane is ruled out, as we require that the temperature at the onset of radiation domination exceeds 3 MeV in order to generate the neutrinos that are needed to produce the observed light element abundances~\cite{KKS99, KKS00, Han04, IKT05}, as well as the observed anisotropies in the cosmic microwave background and large-scale density perturbations~\cite{IKT07,dBPM08}.  

The grey region of Fig.~\ref{plane} is currently ruled out by Fermi's measurement of the high-latitude gamma-ray background, based on our conservative computation of $B_0$. Across this region of parameter space, the EMDE leads to the formation and survival of a large population of microhalos, resulting in large boost factors that are inconsistent with Fermi's observations. Using our conservative $B_0$ values, short EMDEs ($T_\mathrm{dom}/T_\mathrm{RH} \lesssim 10$) do not sufficiently enhance the dark matter annihilation rate to contradict Fermi observations for $m_X \gtrsim 500$ GeV, but these scenarios are ruled out if we use our default $B_0$ computation, as seen in Fig.~\ref{limits}.
In the white region in the lower right portion of this figure, however, gravitational heating suppresses the abundance of microhalos, reducing the boost factors and resulting gamma-ray emission to acceptable levels for both our conservative and default computations.  

\section{Summary and Conclusions}
\label{summary}

If the dark matter is part of a hidden sector that is only very feebly coupled to the Standard Model, the lightest particle within this sector would generically be expected to be long-lived. In such a model, these long-lived particles could come to dominate the energy density of the universe prior to their decays. During such an era of early matter domination, density perturbations would grow more quickly than predicted in standard radiation-dominated cosmology, leading to the formation of a large population of sub-earth-mass dark matter microhalos. 

In this study, we have calculated the evolution of density perturbations in a class of hidden sector dark matter models and estimated how the enhancement of these perturbations during the EMDE affects the microhalo abundance.  The key difference between this analysis and earlier considerations of microhalo formation following an EMDE with weakly interacting dark matter \cite{E15, ESW16} derives from the fact that the dark matter within our hidden sector models does not couple directly to the particles in the Standard Model and is therefore much colder.  As a result, it is possible to form microhalos while the universe is still radiation dominated, and we have employed both spherical collapse theory and $N$-body simulations to determine the formation time and properties of these microhalos.  
We have also studied the case in which the decay of the unstable hidden sector particles occurs after gravitationally bound structures have begun to form.  The disruption of these halos at reheating effectively heats the dark matter and suppresses subsequent microhalo formation. 

The existence of a large microhalo population can cause the dark matter to annihilate at a very high rate, producing detectable fluxes of gamma rays and other annihilation products. Unlike with typical dark matter annihilation, the annihilation rate in a microhalo-dominated scenario scales with the density of dark matter (as opposed to the square of the density), leading to indirect detection signals that resemble those ordinarily predicted from decaying dark matter. Using Fermi's measurement of the high-latitude gamma-ray background, we are able to rule out a wide range of parameter space within this class of models. In particular, if there was an early matter-dominated era that persisted for less than a factor of $\sim\!\! 10^3$ in scale factor, we predict that the dark matter will largely be bound within microhalos, leading to very large annihilation boost factors. In contrast, we find that scenarios with even longer early matter-dominated eras cannot be excluded, as the microhalo abundance is efficiently suppressed in this case by the gravitational heating of the dark matter prior to reheating.


\begin{acknowledgments} 

This manuscript has been authored by Fermi Research Alliance, LLC under Contract No. DE-AC02-07CH11359 with the U.S. Department of Energy, Office of High Energy Physics.  The simulations for this work were carried out on the Dogwood computing cluster at the University of North Carolina at Chapel Hill. M.~S.~D. and A.~L.~E. were partially supported by NSF Grant No. PHY-1752752.

\end{acknowledgments}

\bibliography{EMDEmicrohalos}

\begin{thebibliography}{73}%
\makeatletter
\providecommand \@ifxundefined [1]{%
 \@ifx{#1\undefined}
}%
\providecommand \@ifnum [1]{%
 \ifnum #1\expandafter \@firstoftwo
 \else \expandafter \@secondoftwo
 \fi
}%
\providecommand \@ifx [1]{%
 \ifx #1\expandafter \@firstoftwo
 \else \expandafter \@secondoftwo
 \fi
}%
\providecommand \natexlab [1]{#1}%
\providecommand \enquote  [1]{``#1''}%
\providecommand \bibnamefont  [1]{#1}%
\providecommand \bibfnamefont [1]{#1}%
\providecommand \citenamefont [1]{#1}%
\providecommand \href@noop [0]{\@secondoftwo}%
\providecommand \href [0]{\begingroup \@sanitize@url \@href}%
\providecommand \@href[1]{\@@startlink{#1}\@@href}%
\providecommand \@@href[1]{\endgroup#1\@@endlink}%
\providecommand \@sanitize@url [0]{\catcode `\\12\catcode `\$12\catcode
  `\&12\catcode `\#12\catcode `\^12\catcode `\_12\catcode `\%12\relax}%
\providecommand \@@startlink[1]{}%
\providecommand \@@endlink[0]{}%
\providecommand \url  [0]{\begingroup\@sanitize@url \@url }%
\providecommand \@url [1]{\endgroup\@href {#1}{\urlprefix }}%
\providecommand \urlprefix  [0]{URL }%
\providecommand \Eprint [0]{\href }%
\providecommand \doibase [0]{http://dx.doi.org/}%
\providecommand \selectlanguage [0]{\@gobble}%
\providecommand \bibinfo  [0]{\@secondoftwo}%
\providecommand \bibfield  [0]{\@secondoftwo}%
\providecommand \translation [1]{[#1]}%
\providecommand \BibitemOpen [0]{}%
\providecommand \bibitemStop [0]{}%
\providecommand \bibitemNoStop [0]{.\EOS\space}%
\providecommand \EOS [0]{\spacefactor3000\relax}%
\providecommand \BibitemShut  [1]{\csname bibitem#1\endcsname}%
\let\auto@bib@innerbib\@empty
\bibitem [{\citenamefont {Aprile}\ \emph {et~al.}(2018)\citenamefont {Aprile}
  \emph {et~al.}}]{Aprile:2018dbl}%
  \BibitemOpen
  \bibfield  {author} {\bibinfo {author} {\bibfnamefont {E.}~\bibnamefont
  {Aprile}} \emph {et~al.} (\bibinfo {collaboration} {XENON}),\ }\href
  {\doibase 10.1103/PhysRevLett.121.111302} {\bibfield  {journal} {\bibinfo
  {journal} {Phys. Rev. Lett.}\ }\textbf {\bibinfo {volume} {121}},\ \bibinfo
  {pages} {111302} (\bibinfo {year} {2018})},\ \Eprint
  {http://arxiv.org/abs/1805.12562} {arXiv:1805.12562 [astro-ph.CO]}
  \BibitemShut {NoStop}%
\bibitem [{\citenamefont {Akerib}\ \emph {et~al.}(2017)\citenamefont {Akerib}
  \emph {et~al.}}]{Akerib:2016vxi}%
  \BibitemOpen
  \bibfield  {author} {\bibinfo {author} {\bibfnamefont {D.~S.}\ \bibnamefont
  {Akerib}} \emph {et~al.} (\bibinfo {collaboration} {LUX}),\ }\href {\doibase
  10.1103/PhysRevLett.118.021303} {\bibfield  {journal} {\bibinfo  {journal}
  {Phys. Rev. Lett.}\ }\textbf {\bibinfo {volume} {118}},\ \bibinfo {pages}
  {021303} (\bibinfo {year} {2017})},\ \Eprint
  {http://arxiv.org/abs/1608.07648} {arXiv:1608.07648 [astro-ph.CO]}
  \BibitemShut {NoStop}%
\bibitem [{\citenamefont {Cui}\ \emph {et~al.}(2017)\citenamefont {Cui} \emph
  {et~al.}}]{Cui:2017nnn}%
  \BibitemOpen
  \bibfield  {author} {\bibinfo {author} {\bibfnamefont {X.}~\bibnamefont
  {Cui}} \emph {et~al.} (\bibinfo {collaboration} {PandaX-II}),\ }\href
  {\doibase 10.1103/PhysRevLett.119.181302} {\bibfield  {journal} {\bibinfo
  {journal} {Phys. Rev. Lett.}\ }\textbf {\bibinfo {volume} {119}},\ \bibinfo
  {pages} {181302} (\bibinfo {year} {2017})},\ \Eprint
  {http://arxiv.org/abs/1708.06917} {arXiv:1708.06917 [astro-ph.CO]}
  \BibitemShut {NoStop}%
\bibitem [{\citenamefont {Sirunyan}\ \emph {et~al.}(2019)\citenamefont
  {Sirunyan} \emph {et~al.}}]{Sirunyan:2018dub}%
  \BibitemOpen
  \bibfield  {author} {\bibinfo {author} {\bibfnamefont {A.~M.}\ \bibnamefont
  {Sirunyan}} \emph {et~al.} (\bibinfo {collaboration} {CMS}),\ }\href
  {\doibase 10.1103/PhysRevLett.122.011803} {\bibfield  {journal} {\bibinfo
  {journal} {Phys. Rev. Lett.}\ }\textbf {\bibinfo {volume} {122}},\ \bibinfo
  {pages} {011803} (\bibinfo {year} {2019})},\ \Eprint
  {http://arxiv.org/abs/1807.06522} {arXiv:1807.06522 [hep-ex]} \BibitemShut
  {NoStop}%
\bibitem [{\citenamefont {Sirunyan}\ \emph
  {et~al.}(2018{\natexlab{a}})\citenamefont {Sirunyan} \emph
  {et~al.}}]{Sirunyan:2018xlo}%
  \BibitemOpen
  \bibfield  {author} {\bibinfo {author} {\bibfnamefont {A.~M.}\ \bibnamefont
  {Sirunyan}} \emph {et~al.} (\bibinfo {collaboration} {CMS}),\ }\href
  {\doibase 10.1007/JHEP08(2018)130} {\bibfield  {journal} {\bibinfo  {journal}
  {JHEP}\ }\textbf {\bibinfo {volume} {08}},\ \bibinfo {pages} {130} (\bibinfo
  {year} {2018}{\natexlab{a}})},\ \Eprint {http://arxiv.org/abs/1806.00843}
  {arXiv:1806.00843 [hep-ex]} \BibitemShut {NoStop}%
\bibitem [{\citenamefont {Sirunyan}\ \emph
  {et~al.}(2018{\natexlab{b}})\citenamefont {Sirunyan} \emph
  {et~al.}}]{Sirunyan:2018fpy}%
  \BibitemOpen
  \bibfield  {author} {\bibinfo {author} {\bibfnamefont {A.~M.}\ \bibnamefont
  {Sirunyan}} \emph {et~al.} (\bibinfo {collaboration} {CMS}),\ }\href
  {\doibase 10.1007/JHEP09(2018)046} {\bibfield  {journal} {\bibinfo  {journal}
  {JHEP}\ }\textbf {\bibinfo {volume} {09}},\ \bibinfo {pages} {046} (\bibinfo
  {year} {2018}{\natexlab{b}})},\ \Eprint {http://arxiv.org/abs/1806.04771}
  {arXiv:1806.04771 [hep-ex]} \BibitemShut {NoStop}%
\bibitem [{\citenamefont {Sirunyan}\ \emph
  {et~al.}(2018{\natexlab{c}})\citenamefont {Sirunyan} \emph
  {et~al.}}]{Sirunyan:2018wcm}%
  \BibitemOpen
  \bibfield  {author} {\bibinfo {author} {\bibfnamefont {A.~M.}\ \bibnamefont
  {Sirunyan}} \emph {et~al.} (\bibinfo {collaboration} {CMS}),\ }\href
  {\doibase 10.1140/epjc/s10052-018-6242-x} {\bibfield  {journal} {\bibinfo
  {journal} {Eur. Phys. J.}\ }\textbf {\bibinfo {volume} {C78}},\ \bibinfo
  {pages} {789} (\bibinfo {year} {2018}{\natexlab{c}})},\ \Eprint
  {http://arxiv.org/abs/1803.08030} {arXiv:1803.08030 [hep-ex]} \BibitemShut
  {NoStop}%
\bibitem [{\citenamefont {Sirunyan}\ \emph
  {et~al.}(2018{\natexlab{d}})\citenamefont {Sirunyan} \emph
  {et~al.}}]{Sirunyan:2018gka}%
  \BibitemOpen
  \bibfield  {author} {\bibinfo {author} {\bibfnamefont {A.~M.}\ \bibnamefont
  {Sirunyan}} \emph {et~al.} (\bibinfo {collaboration} {CMS}),\ }\href
  {\doibase 10.1007/JHEP06(2018)027} {\bibfield  {journal} {\bibinfo  {journal}
  {JHEP}\ }\textbf {\bibinfo {volume} {06}},\ \bibinfo {pages} {027} (\bibinfo
  {year} {2018}{\natexlab{d}})},\ \Eprint {http://arxiv.org/abs/1801.08427}
  {arXiv:1801.08427 [hep-ex]} \BibitemShut {NoStop}%
\bibitem [{\citenamefont {Sirunyan}\ \emph
  {et~al.}(2018{\natexlab{e}})\citenamefont {Sirunyan} \emph
  {et~al.}}]{Sirunyan:2017leh}%
  \BibitemOpen
  \bibfield  {author} {\bibinfo {author} {\bibfnamefont {A.~M.}\ \bibnamefont
  {Sirunyan}} \emph {et~al.} (\bibinfo {collaboration} {CMS}),\ }\href
  {\doibase 10.1103/PhysRevD.97.032009} {\bibfield  {journal} {\bibinfo
  {journal} {Phys. Rev.}\ }\textbf {\bibinfo {volume} {D97}},\ \bibinfo {pages}
  {032009} (\bibinfo {year} {2018}{\natexlab{e}})},\ \Eprint
  {http://arxiv.org/abs/1711.00752} {arXiv:1711.00752 [hep-ex]} \BibitemShut
  {NoStop}%
\bibitem [{\citenamefont {Aaboud}\ \emph
  {et~al.}(2018{\natexlab{a}})\citenamefont {Aaboud} \emph
  {et~al.}}]{Aaboud:2018xdl}%
  \BibitemOpen
  \bibfield  {author} {\bibinfo {author} {\bibfnamefont {M.}~\bibnamefont
  {Aaboud}} \emph {et~al.} (\bibinfo {collaboration} {ATLAS}),\ }\href
  {\doibase 10.1007/JHEP10(2018)180} {\bibfield  {journal} {\bibinfo  {journal}
  {JHEP}\ }\textbf {\bibinfo {volume} {10}},\ \bibinfo {pages} {180} (\bibinfo
  {year} {2018}{\natexlab{a}})},\ \Eprint {http://arxiv.org/abs/1807.11471}
  {arXiv:1807.11471 [hep-ex]} \BibitemShut {NoStop}%
\bibitem [{\citenamefont {Aaboud}\ \emph
  {et~al.}(2018{\natexlab{b}})\citenamefont {Aaboud} \emph
  {et~al.}}]{Aaboud:2017phn}%
  \BibitemOpen
  \bibfield  {author} {\bibinfo {author} {\bibfnamefont {M.}~\bibnamefont
  {Aaboud}} \emph {et~al.} (\bibinfo {collaboration} {ATLAS}),\ }\href
  {\doibase 10.1007/JHEP01(2018)126} {\bibfield  {journal} {\bibinfo  {journal}
  {JHEP}\ }\textbf {\bibinfo {volume} {01}},\ \bibinfo {pages} {126} (\bibinfo
  {year} {2018}{\natexlab{b}})},\ \Eprint {http://arxiv.org/abs/1711.03301}
  {arXiv:1711.03301 [hep-ex]} \BibitemShut {NoStop}%
\bibitem [{\citenamefont {Aaboud}\ \emph
  {et~al.}(2018{\natexlab{c}})\citenamefont {Aaboud} \emph
  {et~al.}}]{Aaboud:2017rzf}%
  \BibitemOpen
  \bibfield  {author} {\bibinfo {author} {\bibfnamefont {M.}~\bibnamefont
  {Aaboud}} \emph {et~al.} (\bibinfo {collaboration} {ATLAS}),\ }\href
  {\doibase 10.1140/epjc/s10052-017-5486-1} {\bibfield  {journal} {\bibinfo
  {journal} {Eur. Phys. J.}\ }\textbf {\bibinfo {volume} {C78}},\ \bibinfo
  {pages} {18} (\bibinfo {year} {2018}{\natexlab{c}})},\ \Eprint
  {http://arxiv.org/abs/1710.11412} {arXiv:1710.11412 [hep-ex]} \BibitemShut
  {NoStop}%
\bibitem [{\citenamefont {Aaboud}\ \emph
  {et~al.}(2018{\natexlab{d}})\citenamefont {Aaboud} \emph
  {et~al.}}]{Aaboud:2017bja}%
  \BibitemOpen
  \bibfield  {author} {\bibinfo {author} {\bibfnamefont {M.}~\bibnamefont
  {Aaboud}} \emph {et~al.} (\bibinfo {collaboration} {ATLAS}),\ }\href
  {\doibase 10.1016/j.physletb.2017.11.049} {\bibfield  {journal} {\bibinfo
  {journal} {Phys. Lett.}\ }\textbf {\bibinfo {volume} {B776}},\ \bibinfo
  {pages} {318} (\bibinfo {year} {2018}{\natexlab{d}})},\ \Eprint
  {http://arxiv.org/abs/1708.09624} {arXiv:1708.09624 [hep-ex]} \BibitemShut
  {NoStop}%
\bibitem [{\citenamefont {Aaboud}\ \emph {et~al.}(2017)\citenamefont {Aaboud}
  \emph {et~al.}}]{Aaboud:2017yqz}%
  \BibitemOpen
  \bibfield  {author} {\bibinfo {author} {\bibfnamefont {M.}~\bibnamefont
  {Aaboud}} \emph {et~al.} (\bibinfo {collaboration} {ATLAS}),\ }\href
  {\doibase 10.1103/PhysRevLett.119.181804} {\bibfield  {journal} {\bibinfo
  {journal} {Phys. Rev. Lett.}\ }\textbf {\bibinfo {volume} {119}},\ \bibinfo
  {pages} {181804} (\bibinfo {year} {2017})},\ \Eprint
  {http://arxiv.org/abs/1707.01302} {arXiv:1707.01302 [hep-ex]} \BibitemShut
  {NoStop}%
\bibitem [{\citenamefont {Pospelov}\ \emph {et~al.}(2008)\citenamefont
  {Pospelov}, \citenamefont {Ritz},\ and\ \citenamefont
  {Voloshin}}]{Pospelov:2007mp}%
  \BibitemOpen
  \bibfield  {author} {\bibinfo {author} {\bibfnamefont {M.}~\bibnamefont
  {Pospelov}}, \bibinfo {author} {\bibfnamefont {A.}~\bibnamefont {Ritz}}, \
  and\ \bibinfo {author} {\bibfnamefont {M.~B.}\ \bibnamefont {Voloshin}},\
  }\href {\doibase 10.1016/j.physletb.2008.02.052} {\bibfield  {journal}
  {\bibinfo  {journal} {Phys. Lett.}\ }\textbf {\bibinfo {volume} {B662}},\
  \bibinfo {pages} {53} (\bibinfo {year} {2008})},\ \Eprint
  {http://arxiv.org/abs/0711.4866} {arXiv:0711.4866 [hep-ph]} \BibitemShut
  {NoStop}%
\bibitem [{\citenamefont {Arkani-Hamed}\ \emph {et~al.}(2009)\citenamefont
  {Arkani-Hamed}, \citenamefont {Finkbeiner}, \citenamefont {Slatyer},\ and\
  \citenamefont {Weiner}}]{ArkaniHamed:2008qn}%
  \BibitemOpen
  \bibfield  {author} {\bibinfo {author} {\bibfnamefont {N.}~\bibnamefont
  {Arkani-Hamed}}, \bibinfo {author} {\bibfnamefont {D.~P.}\ \bibnamefont
  {Finkbeiner}}, \bibinfo {author} {\bibfnamefont {T.~R.}\ \bibnamefont
  {Slatyer}}, \ and\ \bibinfo {author} {\bibfnamefont {N.}~\bibnamefont
  {Weiner}},\ }\href {\doibase 10.1103/PhysRevD.79.015014} {\bibfield
  {journal} {\bibinfo  {journal} {Phys. Rev.}\ }\textbf {\bibinfo {volume}
  {D79}},\ \bibinfo {pages} {015014} (\bibinfo {year} {2009})},\ \Eprint
  {http://arxiv.org/abs/0810.0713} {arXiv:0810.0713 [hep-ph]} \BibitemShut
  {NoStop}%
\bibitem [{\citenamefont {Abdullah}\ \emph {et~al.}(2014)\citenamefont
  {Abdullah}, \citenamefont {DiFranzo}, \citenamefont {Rajaraman},
  \citenamefont {Tait}, \citenamefont {Tanedo},\ and\ \citenamefont
  {Wijangco}}]{Abdullah:2014lla}%
  \BibitemOpen
  \bibfield  {author} {\bibinfo {author} {\bibfnamefont {M.}~\bibnamefont
  {Abdullah}}, \bibinfo {author} {\bibfnamefont {A.}~\bibnamefont {DiFranzo}},
  \bibinfo {author} {\bibfnamefont {A.}~\bibnamefont {Rajaraman}}, \bibinfo
  {author} {\bibfnamefont {T.~M.~P.}\ \bibnamefont {Tait}}, \bibinfo {author}
  {\bibfnamefont {P.}~\bibnamefont {Tanedo}}, \ and\ \bibinfo {author}
  {\bibfnamefont {A.~M.}\ \bibnamefont {Wijangco}},\ }\href {\doibase
  10.1103/PhysRevD.90.035004} {\bibfield  {journal} {\bibinfo  {journal} {Phys.
  Rev.}\ }\textbf {\bibinfo {volume} {D90}},\ \bibinfo {pages} {035004}
  (\bibinfo {year} {2014})},\ \Eprint {http://arxiv.org/abs/1404.6528}
  {arXiv:1404.6528 [hep-ph]} \BibitemShut {NoStop}%
\bibitem [{\citenamefont {Berlin}\ \emph {et~al.}(2014)\citenamefont {Berlin},
  \citenamefont {Gratia}, \citenamefont {Hooper},\ and\ \citenamefont
  {McDermott}}]{Berlin:2014pya}%
  \BibitemOpen
  \bibfield  {author} {\bibinfo {author} {\bibfnamefont {A.}~\bibnamefont
  {Berlin}}, \bibinfo {author} {\bibfnamefont {P.}~\bibnamefont {Gratia}},
  \bibinfo {author} {\bibfnamefont {D.}~\bibnamefont {Hooper}}, \ and\ \bibinfo
  {author} {\bibfnamefont {S.~D.}\ \bibnamefont {McDermott}},\ }\href {\doibase
  10.1103/PhysRevD.90.015032} {\bibfield  {journal} {\bibinfo  {journal} {Phys.
  Rev.}\ }\textbf {\bibinfo {volume} {D90}},\ \bibinfo {pages} {015032}
  (\bibinfo {year} {2014})},\ \Eprint {http://arxiv.org/abs/1405.5204}
  {arXiv:1405.5204 [hep-ph]} \BibitemShut {NoStop}%
\bibitem [{\citenamefont {Martin}\ \emph {et~al.}(2014)\citenamefont {Martin},
  \citenamefont {Shelton},\ and\ \citenamefont {Unwin}}]{Martin:2014sxa}%
  \BibitemOpen
  \bibfield  {author} {\bibinfo {author} {\bibfnamefont {A.}~\bibnamefont
  {Martin}}, \bibinfo {author} {\bibfnamefont {J.}~\bibnamefont {Shelton}}, \
  and\ \bibinfo {author} {\bibfnamefont {J.}~\bibnamefont {Unwin}},\ }\href
  {\doibase 10.1103/PhysRevD.90.103513} {\bibfield  {journal} {\bibinfo
  {journal} {Phys. Rev.}\ }\textbf {\bibinfo {volume} {D90}},\ \bibinfo {pages}
  {103513} (\bibinfo {year} {2014})},\ \Eprint {http://arxiv.org/abs/1405.0272}
  {arXiv:1405.0272 [hep-ph]} \BibitemShut {NoStop}%
\bibitem [{\citenamefont {Hooper}\ \emph {et~al.}(2012)\citenamefont {Hooper},
  \citenamefont {Weiner},\ and\ \citenamefont {Xue}}]{Hooper:2012cw}%
  \BibitemOpen
  \bibfield  {author} {\bibinfo {author} {\bibfnamefont {D.}~\bibnamefont
  {Hooper}}, \bibinfo {author} {\bibfnamefont {N.}~\bibnamefont {Weiner}}, \
  and\ \bibinfo {author} {\bibfnamefont {W.}~\bibnamefont {Xue}},\ }\href
  {\doibase 10.1103/PhysRevD.86.056009} {\bibfield  {journal} {\bibinfo
  {journal} {Phys. Rev.}\ }\textbf {\bibinfo {volume} {D86}},\ \bibinfo {pages}
  {056009} (\bibinfo {year} {2012})},\ \Eprint {http://arxiv.org/abs/1206.2929}
  {arXiv:1206.2929 [hep-ph]} \BibitemShut {NoStop}%
\bibitem [{\citenamefont {{Berlin}}\ \emph
  {et~al.}(2016{\natexlab{a}})\citenamefont {{Berlin}}, \citenamefont
  {{Hooper}},\ and\ \citenamefont {{Krnjaic}}}]{BHK16a}%
  \BibitemOpen
  \bibfield  {author} {\bibinfo {author} {\bibfnamefont {A.}~\bibnamefont
  {{Berlin}}}, \bibinfo {author} {\bibfnamefont {D.}~\bibnamefont {{Hooper}}},
  \ and\ \bibinfo {author} {\bibfnamefont {G.}~\bibnamefont {{Krnjaic}}},\
  }\href {\doibase 10.1016/j.physletb.2016.06.037} {\bibfield  {journal}
  {\bibinfo  {journal} {Physics Letters B}\ }\textbf {\bibinfo {volume}
  {760}},\ \bibinfo {pages} {106} (\bibinfo {year} {2016}{\natexlab{a}})},\
  \Eprint {http://arxiv.org/abs/1602.08490} {arXiv:1602.08490 [hep-ph]}
  \BibitemShut {NoStop}%
\bibitem [{\citenamefont {{Berlin}}\ \emph
  {et~al.}(2016{\natexlab{b}})\citenamefont {{Berlin}}, \citenamefont
  {{Hooper}},\ and\ \citenamefont {{Krnjaic}}}]{BHK16b}%
  \BibitemOpen
  \bibfield  {author} {\bibinfo {author} {\bibfnamefont {A.}~\bibnamefont
  {{Berlin}}}, \bibinfo {author} {\bibfnamefont {D.}~\bibnamefont {{Hooper}}},
  \ and\ \bibinfo {author} {\bibfnamefont {G.}~\bibnamefont {{Krnjaic}}},\
  }\href {\doibase 10.1103/PhysRevD.94.095019} {\bibfield  {journal} {\bibinfo
  {journal} {\prd}\ }\textbf {\bibinfo {volume} {94}},\ \bibinfo {eid} {095019}
  (\bibinfo {year} {2016}{\natexlab{b}})},\ \Eprint
  {http://arxiv.org/abs/1609.02555} {arXiv:1609.02555 [hep-ph]} \BibitemShut
  {NoStop}%
\bibitem [{\citenamefont {Dror}\ \emph {et~al.}(2016)\citenamefont {Dror},
  \citenamefont {Kuflik},\ and\ \citenamefont {Ng}}]{Dror:2016rxc}%
  \BibitemOpen
  \bibfield  {author} {\bibinfo {author} {\bibfnamefont {J.~A.}\ \bibnamefont
  {Dror}}, \bibinfo {author} {\bibfnamefont {E.}~\bibnamefont {Kuflik}}, \ and\
  \bibinfo {author} {\bibfnamefont {W.~H.}\ \bibnamefont {Ng}},\ }\href
  {\doibase 10.1103/PhysRevLett.117.211801} {\bibfield  {journal} {\bibinfo
  {journal} {Phys. Rev. Lett.}\ }\textbf {\bibinfo {volume} {117}},\ \bibinfo
  {pages} {211801} (\bibinfo {year} {2016})},\ \Eprint
  {http://arxiv.org/abs/1607.03110} {arXiv:1607.03110 [hep-ph]} \BibitemShut
  {NoStop}%
\bibitem [{\citenamefont {Dror}\ \emph {et~al.}(2018)\citenamefont {Dror},
  \citenamefont {Kuflik}, \citenamefont {Melcher},\ and\ \citenamefont
  {Watson}}]{Dror:2017gjq}%
  \BibitemOpen
  \bibfield  {author} {\bibinfo {author} {\bibfnamefont {J.~A.}\ \bibnamefont
  {Dror}}, \bibinfo {author} {\bibfnamefont {E.}~\bibnamefont {Kuflik}},
  \bibinfo {author} {\bibfnamefont {B.}~\bibnamefont {Melcher}}, \ and\
  \bibinfo {author} {\bibfnamefont {S.}~\bibnamefont {Watson}},\ }\href
  {\doibase 10.1103/PhysRevD.97.063524} {\bibfield  {journal} {\bibinfo
  {journal} {Phys. Rev.}\ }\textbf {\bibinfo {volume} {D97}},\ \bibinfo {pages}
  {063524} (\bibinfo {year} {2018})},\ \Eprint
  {http://arxiv.org/abs/1711.04773} {arXiv:1711.04773 [hep-ph]} \BibitemShut
  {NoStop}%
\bibitem [{\citenamefont {Steigman}\ \emph {et~al.}(2012)\citenamefont
  {Steigman}, \citenamefont {Dasgupta},\ and\ \citenamefont
  {Beacom}}]{Steigman:2012nb}%
  \BibitemOpen
  \bibfield  {author} {\bibinfo {author} {\bibfnamefont {G.}~\bibnamefont
  {Steigman}}, \bibinfo {author} {\bibfnamefont {B.}~\bibnamefont {Dasgupta}},
  \ and\ \bibinfo {author} {\bibfnamefont {J.~F.}\ \bibnamefont {Beacom}},\
  }\href {\doibase 10.1103/PhysRevD.86.023506} {\bibfield  {journal} {\bibinfo
  {journal} {Phys. Rev.}\ }\textbf {\bibinfo {volume} {D86}},\ \bibinfo {pages}
  {023506} (\bibinfo {year} {2012})},\ \Eprint {http://arxiv.org/abs/1204.3622}
  {arXiv:1204.3622 [hep-ph]} \BibitemShut {NoStop}%
\bibitem [{\citenamefont {Albert}\ \emph {et~al.}(2017)\citenamefont {Albert}
  \emph {et~al.}}]{Fermi-LAT:2016uux}%
  \BibitemOpen
  \bibfield  {author} {\bibinfo {author} {\bibfnamefont {A.}~\bibnamefont
  {Albert}} \emph {et~al.} (\bibinfo {collaboration} {Fermi-LAT, DES}),\ }\href
  {\doibase 10.3847/1538-4357/834/2/110} {\bibfield  {journal} {\bibinfo
  {journal} {Astrophys. J.}\ }\textbf {\bibinfo {volume} {834}},\ \bibinfo
  {pages} {110} (\bibinfo {year} {2017})},\ \Eprint
  {http://arxiv.org/abs/1611.03184} {arXiv:1611.03184 [astro-ph.HE]}
  \BibitemShut {NoStop}%
\bibitem [{\citenamefont {Cholis}\ \emph {et~al.}(2019)\citenamefont {Cholis},
  \citenamefont {Linden},\ and\ \citenamefont {Hooper}}]{Cholis:2019ejx}%
  \BibitemOpen
  \bibfield  {author} {\bibinfo {author} {\bibfnamefont {I.}~\bibnamefont
  {Cholis}}, \bibinfo {author} {\bibfnamefont {T.}~\bibnamefont {Linden}}, \
  and\ \bibinfo {author} {\bibfnamefont {D.}~\bibnamefont {Hooper}},\
  }\href@noop {} {\  (\bibinfo {year} {2019})},\ \Eprint
  {http://arxiv.org/abs/1903.02549} {arXiv:1903.02549 [astro-ph.HE]}
  \BibitemShut {NoStop}%
\bibitem [{\citenamefont {Bergstrom}\ \emph {et~al.}(2013)\citenamefont
  {Bergstrom}, \citenamefont {Bringmann}, \citenamefont {Cholis}, \citenamefont
  {Hooper},\ and\ \citenamefont {Weniger}}]{Bergstrom:2013jra}%
  \BibitemOpen
  \bibfield  {author} {\bibinfo {author} {\bibfnamefont {L.}~\bibnamefont
  {Bergstrom}}, \bibinfo {author} {\bibfnamefont {T.}~\bibnamefont
  {Bringmann}}, \bibinfo {author} {\bibfnamefont {I.}~\bibnamefont {Cholis}},
  \bibinfo {author} {\bibfnamefont {D.}~\bibnamefont {Hooper}}, \ and\ \bibinfo
  {author} {\bibfnamefont {C.}~\bibnamefont {Weniger}},\ }\href {\doibase
  10.1103/PhysRevLett.111.171101} {\bibfield  {journal} {\bibinfo  {journal}
  {Phys.Rev.Lett.}\ }\textbf {\bibinfo {volume} {111}},\ \bibinfo {pages}
  {171101} (\bibinfo {year} {2013})},\ \Eprint {http://arxiv.org/abs/1306.3983}
  {arXiv:1306.3983 [astro-ph.HE]} \BibitemShut {NoStop}%
\bibitem [{\citenamefont {{Erickcek}}\ and\ \citenamefont
  {{Sigurdson}}(2011)}]{ES11}%
  \BibitemOpen
  \bibfield  {author} {\bibinfo {author} {\bibfnamefont {A.~L.}\ \bibnamefont
  {{Erickcek}}}\ and\ \bibinfo {author} {\bibfnamefont {K.}~\bibnamefont
  {{Sigurdson}}},\ }\href {\doibase 10.1103/PhysRevD.84.083503} {\bibfield
  {journal} {\bibinfo  {journal} {\prd}\ }\textbf {\bibinfo {volume} {84}},\
  \bibinfo {eid} {083503} (\bibinfo {year} {2011})},\ \Eprint
  {http://arxiv.org/abs/1106.0536} {arXiv:1106.0536 [astro-ph.CO]} \BibitemShut
  {NoStop}%
\bibitem [{\citenamefont {{Barenboim}}\ and\ \citenamefont
  {{Rasero}}(2014)}]{BR14}%
  \BibitemOpen
  \bibfield  {author} {\bibinfo {author} {\bibfnamefont {G.}~\bibnamefont
  {{Barenboim}}}\ and\ \bibinfo {author} {\bibfnamefont {J.}~\bibnamefont
  {{Rasero}}},\ }\href {\doibase 10.1007/JHEP04(2014)138} {\bibfield  {journal}
  {\bibinfo  {journal} {Journal of High Energy Physics}\ }\textbf {\bibinfo
  {volume} {2014}},\ \bibinfo {eid} {138} (\bibinfo {year} {2014})},\ \Eprint
  {http://arxiv.org/abs/1311.4034} {arXiv:1311.4034 [hep-ph]} \BibitemShut
  {NoStop}%
\bibitem [{\citenamefont {{Fan}}\ \emph {et~al.}(2014)\citenamefont {{Fan}},
  \citenamefont {{{\"O}zsoy}},\ and\ \citenamefont {{Watson}}}]{FOW14}%
  \BibitemOpen
  \bibfield  {author} {\bibinfo {author} {\bibfnamefont {J.}~\bibnamefont
  {{Fan}}}, \bibinfo {author} {\bibfnamefont {O.}~\bibnamefont {{{\"O}zsoy}}},
  \ and\ \bibinfo {author} {\bibfnamefont {S.}~\bibnamefont {{Watson}}},\
  }\href {\doibase 10.1103/PhysRevD.90.043536} {\bibfield  {journal} {\bibinfo
  {journal} {\prd}\ }\textbf {\bibinfo {volume} {90}},\ \bibinfo {eid} {043536}
  (\bibinfo {year} {2014})},\ \Eprint {http://arxiv.org/abs/1405.7373}
  {arXiv:1405.7373 [hep-ph]} \BibitemShut {NoStop}%
\bibitem [{\citenamefont {{Erickcek}}(2015)}]{E15}%
  \BibitemOpen
  \bibfield  {author} {\bibinfo {author} {\bibfnamefont {A.~L.}\ \bibnamefont
  {{Erickcek}}},\ }\href {\doibase 10.1103/PhysRevD.92.103505} {\bibfield
  {journal} {\bibinfo  {journal} {\prd}\ }\textbf {\bibinfo {volume} {92}},\
  \bibinfo {eid} {103505} (\bibinfo {year} {2015})},\ \Eprint
  {http://arxiv.org/abs/1504.03335} {arXiv:1504.03335} \BibitemShut {NoStop}%
\bibitem [{\citenamefont {{Erickcek}}\ \emph {et~al.}(2016)\citenamefont
  {{Erickcek}}, \citenamefont {{Sinha}},\ and\ \citenamefont
  {{Watson}}}]{ESW16}%
  \BibitemOpen
  \bibfield  {author} {\bibinfo {author} {\bibfnamefont {A.~L.}\ \bibnamefont
  {{Erickcek}}}, \bibinfo {author} {\bibfnamefont {K.}~\bibnamefont {{Sinha}}},
  \ and\ \bibinfo {author} {\bibfnamefont {S.}~\bibnamefont {{Watson}}},\
  }\href {\doibase 10.1103/PhysRevD.94.063502} {\bibfield  {journal} {\bibinfo
  {journal} {\prd}\ }\textbf {\bibinfo {volume} {94}},\ \bibinfo {eid} {063502}
  (\bibinfo {year} {2016})},\ \Eprint {http://arxiv.org/abs/1510.04291}
  {arXiv:1510.04291 [hep-ph]} \BibitemShut {NoStop}%
\bibitem [{\citenamefont {Ackermann}\ \emph {et~al.}(2015)\citenamefont
  {Ackermann} \emph {et~al.}}]{Ackermann:2014usa}%
  \BibitemOpen
  \bibfield  {author} {\bibinfo {author} {\bibfnamefont {M.}~\bibnamefont
  {Ackermann}} \emph {et~al.} (\bibinfo {collaboration} {Fermi-LAT}),\ }\href
  {\doibase 10.1088/0004-637X/799/1/86} {\bibfield  {journal} {\bibinfo
  {journal} {Astrophys. J.}\ }\textbf {\bibinfo {volume} {799}},\ \bibinfo
  {pages} {86} (\bibinfo {year} {2015})},\ \Eprint
  {http://arxiv.org/abs/1410.3696} {arXiv:1410.3696 [astro-ph.HE]} \BibitemShut
  {NoStop}%
\bibitem [{\citenamefont {Blanco}\ and\ \citenamefont
  {Hooper}(2019)}]{Blanco:2018esa}%
  \BibitemOpen
  \bibfield  {author} {\bibinfo {author} {\bibfnamefont {C.}~\bibnamefont
  {Blanco}}\ and\ \bibinfo {author} {\bibfnamefont {D.}~\bibnamefont
  {Hooper}},\ }\href {\doibase 10.1088/1475-7516/2019/03/019} {\bibfield
  {journal} {\bibinfo  {journal} {JCAP}\ }\textbf {\bibinfo {volume} {1903}},\
  \bibinfo {pages} {019} (\bibinfo {year} {2019})},\ \Eprint
  {http://arxiv.org/abs/1811.05988} {arXiv:1811.05988 [astro-ph.HE]}
  \BibitemShut {NoStop}%
\bibitem [{\citenamefont {Krolikowski}(2008)}]{Krolikowski:2008qa}%
  \BibitemOpen
  \bibfield  {author} {\bibinfo {author} {\bibfnamefont {W.}~\bibnamefont
  {Krolikowski}},\ }\href@noop {} {\  (\bibinfo {year} {2008})},\ \Eprint
  {http://arxiv.org/abs/0803.2977} {arXiv:0803.2977 [hep-ph]} \BibitemShut
  {NoStop}%
\bibitem [{\citenamefont {Burgess}\ \emph {et~al.}(2001)\citenamefont
  {Burgess}, \citenamefont {Pospelov},\ and\ \citenamefont {ter
  Veldhuis}}]{Burgess:2000yq}%
  \BibitemOpen
  \bibfield  {author} {\bibinfo {author} {\bibfnamefont {C.~P.}\ \bibnamefont
  {Burgess}}, \bibinfo {author} {\bibfnamefont {M.}~\bibnamefont {Pospelov}}, \
  and\ \bibinfo {author} {\bibfnamefont {T.}~\bibnamefont {ter Veldhuis}},\
  }\href {\doibase 10.1016/S0550-3213(01)00513-2} {\bibfield  {journal}
  {\bibinfo  {journal} {Nucl. Phys.}\ }\textbf {\bibinfo {volume} {B619}},\
  \bibinfo {pages} {709} (\bibinfo {year} {2001})},\ \Eprint
  {http://arxiv.org/abs/hep-ph/0011335} {arXiv:hep-ph/0011335 [hep-ph]}
  \BibitemShut {NoStop}%
\bibitem [{\citenamefont {Davoudiasl}\ \emph {et~al.}(2005)\citenamefont
  {Davoudiasl}, \citenamefont {Kitano}, \citenamefont {Li},\ and\ \citenamefont
  {Murayama}}]{Davoudiasl:2004be}%
  \BibitemOpen
  \bibfield  {author} {\bibinfo {author} {\bibfnamefont {H.}~\bibnamefont
  {Davoudiasl}}, \bibinfo {author} {\bibfnamefont {R.}~\bibnamefont {Kitano}},
  \bibinfo {author} {\bibfnamefont {T.}~\bibnamefont {Li}}, \ and\ \bibinfo
  {author} {\bibfnamefont {H.}~\bibnamefont {Murayama}},\ }\href {\doibase
  10.1016/j.physletb.2005.01.026} {\bibfield  {journal} {\bibinfo  {journal}
  {Phys. Lett.}\ }\textbf {\bibinfo {volume} {B609}},\ \bibinfo {pages} {117}
  (\bibinfo {year} {2005})},\ \Eprint {http://arxiv.org/abs/hep-ph/0405097}
  {arXiv:hep-ph/0405097 [hep-ph]} \BibitemShut {NoStop}%
\bibitem [{\citenamefont {Bird}\ \emph {et~al.}(2006)\citenamefont {Bird},
  \citenamefont {Kowalewski},\ and\ \citenamefont {Pospelov}}]{Bird:2006jd}%
  \BibitemOpen
  \bibfield  {author} {\bibinfo {author} {\bibfnamefont {C.}~\bibnamefont
  {Bird}}, \bibinfo {author} {\bibfnamefont {R.~V.}\ \bibnamefont
  {Kowalewski}}, \ and\ \bibinfo {author} {\bibfnamefont {M.}~\bibnamefont
  {Pospelov}},\ }\href {\doibase 10.1142/S0217732306019852} {\bibfield
  {journal} {\bibinfo  {journal} {Mod. Phys. Lett.}\ }\textbf {\bibinfo
  {volume} {A21}},\ \bibinfo {pages} {457} (\bibinfo {year} {2006})},\ \Eprint
  {http://arxiv.org/abs/hep-ph/0601090} {arXiv:hep-ph/0601090 [hep-ph]}
  \BibitemShut {NoStop}%
\bibitem [{\citenamefont {Kim}\ and\ \citenamefont {Lee}(2007)}]{Kim:2006af}%
  \BibitemOpen
  \bibfield  {author} {\bibinfo {author} {\bibfnamefont {Y.~G.}\ \bibnamefont
  {Kim}}\ and\ \bibinfo {author} {\bibfnamefont {K.~Y.}\ \bibnamefont {Lee}},\
  }\href {\doibase 10.1103/PhysRevD.75.115012} {\bibfield  {journal} {\bibinfo
  {journal} {Phys. Rev.}\ }\textbf {\bibinfo {volume} {D75}},\ \bibinfo {pages}
  {115012} (\bibinfo {year} {2007})},\ \Eprint
  {http://arxiv.org/abs/hep-ph/0611069} {arXiv:hep-ph/0611069 [hep-ph]}
  \BibitemShut {NoStop}%
\bibitem [{\citenamefont {Finkbeiner}\ and\ \citenamefont
  {Weiner}(2007)}]{Finkbeiner:2007kk}%
  \BibitemOpen
  \bibfield  {author} {\bibinfo {author} {\bibfnamefont {D.~P.}\ \bibnamefont
  {Finkbeiner}}\ and\ \bibinfo {author} {\bibfnamefont {N.}~\bibnamefont
  {Weiner}},\ }\href {\doibase 10.1103/PhysRevD.76.083519} {\bibfield
  {journal} {\bibinfo  {journal} {Phys. Rev.}\ }\textbf {\bibinfo {volume}
  {D76}},\ \bibinfo {pages} {083519} (\bibinfo {year} {2007})},\ \Eprint
  {http://arxiv.org/abs/astro-ph/0702587} {arXiv:astro-ph/0702587 [astro-ph]}
  \BibitemShut {NoStop}%
\bibitem [{\citenamefont {D'Eramo}(2007)}]{DEramo:2007anh}%
  \BibitemOpen
  \bibfield  {author} {\bibinfo {author} {\bibfnamefont {F.}~\bibnamefont
  {D'Eramo}},\ }\href {\doibase 10.1103/PhysRevD.76.083522} {\bibfield
  {journal} {\bibinfo  {journal} {Phys. Rev.}\ }\textbf {\bibinfo {volume}
  {D76}},\ \bibinfo {pages} {083522} (\bibinfo {year} {2007})},\ \Eprint
  {http://arxiv.org/abs/0705.4493} {arXiv:0705.4493 [hep-ph]} \BibitemShut
  {NoStop}%
\bibitem [{\citenamefont {March-Russell}\ \emph {et~al.}(2008)\citenamefont
  {March-Russell}, \citenamefont {West}, \citenamefont {Cumberbatch},\ and\
  \citenamefont {Hooper}}]{MarchRussell:2008yu}%
  \BibitemOpen
  \bibfield  {author} {\bibinfo {author} {\bibfnamefont {J.}~\bibnamefont
  {March-Russell}}, \bibinfo {author} {\bibfnamefont {S.~M.}\ \bibnamefont
  {West}}, \bibinfo {author} {\bibfnamefont {D.}~\bibnamefont {Cumberbatch}}, \
  and\ \bibinfo {author} {\bibfnamefont {D.}~\bibnamefont {Hooper}},\ }\href
  {\doibase 10.1088/1126-6708/2008/07/058} {\bibfield  {journal} {\bibinfo
  {journal} {JHEP}\ }\textbf {\bibinfo {volume} {07}},\ \bibinfo {pages} {058}
  (\bibinfo {year} {2008})},\ \Eprint {http://arxiv.org/abs/0801.3440}
  {arXiv:0801.3440 [hep-ph]} \BibitemShut {NoStop}%
\bibitem [{\citenamefont {Bai}\ and\ \citenamefont
  {Berger}(2014)}]{Bai:2014osa}%
  \BibitemOpen
  \bibfield  {author} {\bibinfo {author} {\bibfnamefont {Y.}~\bibnamefont
  {Bai}}\ and\ \bibinfo {author} {\bibfnamefont {J.}~\bibnamefont {Berger}},\
  }\href {\doibase 10.1007/JHEP08(2014)153} {\bibfield  {journal} {\bibinfo
  {journal} {JHEP}\ }\textbf {\bibinfo {volume} {08}},\ \bibinfo {pages} {153}
  (\bibinfo {year} {2014})},\ \Eprint {http://arxiv.org/abs/1402.6696}
  {arXiv:1402.6696 [hep-ph]} \BibitemShut {NoStop}%
\bibitem [{\citenamefont {Holdom}(1986)}]{Holdom:1985ag}%
  \BibitemOpen
  \bibfield  {author} {\bibinfo {author} {\bibfnamefont {B.}~\bibnamefont
  {Holdom}},\ }\href {\doibase 10.1016/0370-2693(86)91377-8} {\bibfield
  {journal} {\bibinfo  {journal} {Phys. Lett.}\ }\textbf {\bibinfo {volume}
  {166B}},\ \bibinfo {pages} {196} (\bibinfo {year} {1986})}\BibitemShut
  {NoStop}%
\bibitem [{\citenamefont {Kolb}\ and\ \citenamefont
  {Turner}(1990)}]{Kolb:1990vq}%
  \BibitemOpen
  \bibfield  {author} {\bibinfo {author} {\bibfnamefont {E.~W.}\ \bibnamefont
  {Kolb}}\ and\ \bibinfo {author} {\bibfnamefont {M.~S.}\ \bibnamefont
  {Turner}},\ }\href@noop {} {\bibfield  {journal} {\bibinfo  {journal} {Front.
  Phys.}\ }\textbf {\bibinfo {volume} {69}},\ \bibinfo {pages} {1} (\bibinfo
  {year} {1990})}\BibitemShut {NoStop}%
\bibitem [{\citenamefont {{Bertschinger}}(2006)}]{Bertschinger06}%
  \BibitemOpen
  \bibfield  {author} {\bibinfo {author} {\bibfnamefont {E.}~\bibnamefont
  {{Bertschinger}}},\ }\href {\doibase 10.1103/PhysRevD.74.063509} {\bibfield
  {journal} {\bibinfo  {journal} {\prd}\ }\textbf {\bibinfo {volume} {74}},\
  \bibinfo {pages} {063509} (\bibinfo {year} {2006})},\ \Eprint
  {http://arxiv.org/abs/arXiv:astro-ph/0607319} {arXiv:astro-ph/0607319}
  \BibitemShut {NoStop}%
\bibitem [{\citenamefont {{Hu}}\ and\ \citenamefont {{Sugiyama}}(1996)}]{HS96}%
  \BibitemOpen
  \bibfield  {author} {\bibinfo {author} {\bibfnamefont {W.}~\bibnamefont
  {{Hu}}}\ and\ \bibinfo {author} {\bibfnamefont {N.}~\bibnamefont
  {{Sugiyama}}},\ }\href {\doibase 10.1086/177989} {\bibfield  {journal}
  {\bibinfo  {journal} {\apj}\ }\textbf {\bibinfo {volume} {471}},\ \bibinfo
  {pages} {542} (\bibinfo {year} {1996})},\ \Eprint
  {http://arxiv.org/abs/arXiv:astro-ph/9510117} {arXiv:astro-ph/9510117}
  \BibitemShut {NoStop}%
\bibitem [{\citenamefont {{Peebles}}(1980)}]{Peebles80}%
  \BibitemOpen
  \bibfield  {author} {\bibinfo {author} {\bibfnamefont {P.~J.~E.}\
  \bibnamefont {{Peebles}}},\ }\href@noop {} {\emph {\bibinfo {title} {{The
  large-scale structure of the universe}}}}\ (\bibinfo  {publisher} {Princeton;
  Princeton University Press},\ \bibinfo {year} {1980})\BibitemShut {NoStop}%
\bibitem [{\citenamefont {{Padmanabhan}}\ and\ \citenamefont
  {{Subramanian}}(1993)}]{PS93}%
  \BibitemOpen
  \bibfield  {author} {\bibinfo {author} {\bibfnamefont {T.}~\bibnamefont
  {{Padmanabhan}}}\ and\ \bibinfo {author} {\bibfnamefont {K.}~\bibnamefont
  {{Subramanian}}},\ }\href {\doibase 10.1086/173286} {\bibfield  {journal}
  {\bibinfo  {journal} {\apj}\ }\textbf {\bibinfo {volume} {417}},\ \bibinfo
  {pages} {3} (\bibinfo {year} {1993})}\BibitemShut {NoStop}%
\bibitem [{\citenamefont {{Zel'dovich}}(1970)}]{Zeldovich70}%
  \BibitemOpen
  \bibfield  {author} {\bibinfo {author} {\bibfnamefont {Y.~B.}\ \bibnamefont
  {{Zel'dovich}}},\ }\href@noop {} {\bibfield  {journal} {\bibinfo  {journal}
  {\aap}\ }\textbf {\bibinfo {volume} {5}},\ \bibinfo {pages} {84} (\bibinfo
  {year} {1970})}\BibitemShut {NoStop}%
\bibitem [{\citenamefont {{Springel}}(2005)}]{2005MNRAS.364.1105S}%
  \BibitemOpen
  \bibfield  {author} {\bibinfo {author} {\bibfnamefont {V.}~\bibnamefont
  {{Springel}}},\ }\href {\doibase 10.1111/j.1365-2966.2005.09655.x} {\bibfield
   {journal} {\bibinfo  {journal} {\mnras}\ }\textbf {\bibinfo {volume}
  {364}},\ \bibinfo {pages} {1105} (\bibinfo {year} {2005})},\ \Eprint
  {http://arxiv.org/abs/astro-ph/0505010} {astro-ph/0505010} \BibitemShut
  {NoStop}%
\bibitem [{\citenamefont {{Delos}}\ \emph
  {et~al.}(2018{\natexlab{a}})\citenamefont {{Delos}}, \citenamefont
  {{Erickcek}}, \citenamefont {{Bailey}},\ and\ \citenamefont
  {{Alvarez}}}]{DEAB18a}%
  \BibitemOpen
  \bibfield  {author} {\bibinfo {author} {\bibfnamefont {M.~S.}\ \bibnamefont
  {{Delos}}}, \bibinfo {author} {\bibfnamefont {A.~L.}\ \bibnamefont
  {{Erickcek}}}, \bibinfo {author} {\bibfnamefont {A.~P.}\ \bibnamefont
  {{Bailey}}}, \ and\ \bibinfo {author} {\bibfnamefont {M.~A.}\ \bibnamefont
  {{Alvarez}}},\ }\href {\doibase 10.1103/PhysRevD.97.041303} {\bibfield
  {journal} {\bibinfo  {journal} {\prd}\ }\textbf {\bibinfo {volume} {97}},\
  \bibinfo {eid} {041303} (\bibinfo {year} {2018}{\natexlab{a}})}\BibitemShut
  {NoStop}%
\bibitem [{\citenamefont {{Delos}}\ \emph
  {et~al.}(2018{\natexlab{b}})\citenamefont {{Delos}}, \citenamefont
  {{Erickcek}}, \citenamefont {{Bailey}},\ and\ \citenamefont
  {{Alvarez}}}]{DEAB18b}%
  \BibitemOpen
  \bibfield  {author} {\bibinfo {author} {\bibfnamefont {M.~S.}\ \bibnamefont
  {{Delos}}}, \bibinfo {author} {\bibfnamefont {A.~L.}\ \bibnamefont
  {{Erickcek}}}, \bibinfo {author} {\bibfnamefont {A.~P.}\ \bibnamefont
  {{Bailey}}}, \ and\ \bibinfo {author} {\bibfnamefont {M.~A.}\ \bibnamefont
  {{Alvarez}}},\ }\href {\doibase 10.1103/PhysRevD.98.063527} {\bibfield
  {journal} {\bibinfo  {journal} {\prd}\ }\textbf {\bibinfo {volume} {98}},\
  \bibinfo {eid} {063527} (\bibinfo {year} {2018}{\natexlab{b}})}\BibitemShut
  {NoStop}%
\bibitem [{\citenamefont {{Suto}}(1987)}]{1987ApJ...321...36S}%
  \BibitemOpen
  \bibfield  {author} {\bibinfo {author} {\bibfnamefont {Y.}~\bibnamefont
  {{Suto}}},\ }\href {\doibase 10.1086/165614} {\bibfield  {journal} {\bibinfo
  {journal} {\apj}\ }\textbf {\bibinfo {volume} {321}},\ \bibinfo {pages} {36}
  (\bibinfo {year} {1987})}\BibitemShut {NoStop}%
\bibitem [{\citenamefont {Enqvist}\ \emph {et~al.}(2015)\citenamefont
  {Enqvist}, \citenamefont {Nadathur}, \citenamefont {Sekiguchi},\ and\
  \citenamefont {Takahashi}}]{Enqvist:2015ara}%
  \BibitemOpen
  \bibfield  {author} {\bibinfo {author} {\bibfnamefont {K.}~\bibnamefont
  {Enqvist}}, \bibinfo {author} {\bibfnamefont {S.}~\bibnamefont {Nadathur}},
  \bibinfo {author} {\bibfnamefont {T.}~\bibnamefont {Sekiguchi}}, \ and\
  \bibinfo {author} {\bibfnamefont {T.}~\bibnamefont {Takahashi}},\ }\href
  {\doibase 10.1088/1475-7516/2015/09/067} {\bibfield  {journal} {\bibinfo
  {journal} {JCAP}\ }\textbf {\bibinfo {volume} {1509}},\ \bibinfo {pages}
  {067} (\bibinfo {year} {2015})},\ \Eprint {http://arxiv.org/abs/1505.05511}
  {arXiv:1505.05511 [astro-ph.CO]} \BibitemShut {NoStop}%
\bibitem [{\citenamefont {Dakin}\ \emph {et~al.}(2019)\citenamefont {Dakin},
  \citenamefont {Hannestad},\ and\ \citenamefont {Tram}}]{Dakin:2019dxu}%
  \BibitemOpen
  \bibfield  {author} {\bibinfo {author} {\bibfnamefont {J.}~\bibnamefont
  {Dakin}}, \bibinfo {author} {\bibfnamefont {S.}~\bibnamefont {Hannestad}}, \
  and\ \bibinfo {author} {\bibfnamefont {T.}~\bibnamefont {Tram}},\ }\href@noop
  {} {\  (\bibinfo {year} {2019})},\ \Eprint {http://arxiv.org/abs/1904.11773}
  {arXiv:1904.11773 [astro-ph.CO]} \BibitemShut {NoStop}%
\bibitem [{\citenamefont {{Bardeen}}\ \emph {et~al.}(1986)\citenamefont
  {{Bardeen}}, \citenamefont {{Bond}}, \citenamefont {{Kaiser}},\ and\
  \citenamefont {{Szalay}}}]{BBKS86}%
  \BibitemOpen
  \bibfield  {author} {\bibinfo {author} {\bibfnamefont {J.~M.}\ \bibnamefont
  {{Bardeen}}}, \bibinfo {author} {\bibfnamefont {J.~R.}\ \bibnamefont
  {{Bond}}}, \bibinfo {author} {\bibfnamefont {N.}~\bibnamefont {{Kaiser}}}, \
  and\ \bibinfo {author} {\bibfnamefont {A.~S.}\ \bibnamefont {{Szalay}}},\
  }\href {\doibase 10.1086/164143} {\bibfield  {journal} {\bibinfo  {journal}
  {\apj}\ }\textbf {\bibinfo {volume} {304}},\ \bibinfo {pages} {15} (\bibinfo
  {year} {1986})}\BibitemShut {NoStop}%
\bibitem [{\citenamefont {{Navarro}}\ \emph {et~al.}(1997)\citenamefont
  {{Navarro}}, \citenamefont {{Frenk}},\ and\ \citenamefont {{White}}}]{NFW97}%
  \BibitemOpen
  \bibfield  {author} {\bibinfo {author} {\bibfnamefont {J.~F.}\ \bibnamefont
  {{Navarro}}}, \bibinfo {author} {\bibfnamefont {C.~S.}\ \bibnamefont
  {{Frenk}}}, \ and\ \bibinfo {author} {\bibfnamefont {S.~D.~M.}\ \bibnamefont
  {{White}}},\ }\href {\doibase 10.1086/304888} {\bibfield  {journal} {\bibinfo
   {journal} {\apj}\ }\textbf {\bibinfo {volume} {490}},\ \bibinfo {pages}
  {493} (\bibinfo {year} {1997})},\ \Eprint
  {http://arxiv.org/abs/astro-ph/9611107} {arXiv:astro-ph/9611107 [astro-ph]}
  \BibitemShut {NoStop}%
\bibitem [{\citenamefont {{Press}}\ and\ \citenamefont
  {{Schechter}}(1974)}]{PS74}%
  \BibitemOpen
  \bibfield  {author} {\bibinfo {author} {\bibfnamefont {W.~H.}\ \bibnamefont
  {{Press}}}\ and\ \bibinfo {author} {\bibfnamefont {P.}~\bibnamefont
  {{Schechter}}},\ }\href {\doibase 10.1086/152650} {\bibfield  {journal}
  {\bibinfo  {journal} {\apj}\ }\textbf {\bibinfo {volume} {187}},\ \bibinfo
  {pages} {425} (\bibinfo {year} {1974})}\BibitemShut {NoStop}%
\bibitem [{\citenamefont {{Benson}}\ \emph {et~al.}(2013)\citenamefont
  {{Benson}}, \citenamefont {{Farahi}}, \citenamefont {{Cole}}, \citenamefont
  {{Moustakas}}, \citenamefont {{Jenkins}}, \citenamefont {{Lovell}},
  \citenamefont {{Kennedy}}, \citenamefont {{Helly}},\ and\ \citenamefont
  {{Frenk}}}]{2013MNRAS.428.1774B}%
  \BibitemOpen
  \bibfield  {author} {\bibinfo {author} {\bibfnamefont {A.~J.}\ \bibnamefont
  {{Benson}}}, \bibinfo {author} {\bibfnamefont {A.}~\bibnamefont {{Farahi}}},
  \bibinfo {author} {\bibfnamefont {S.}~\bibnamefont {{Cole}}}, \bibinfo
  {author} {\bibfnamefont {L.~A.}\ \bibnamefont {{Moustakas}}}, \bibinfo
  {author} {\bibfnamefont {A.}~\bibnamefont {{Jenkins}}}, \bibinfo {author}
  {\bibfnamefont {M.}~\bibnamefont {{Lovell}}}, \bibinfo {author}
  {\bibfnamefont {R.}~\bibnamefont {{Kennedy}}}, \bibinfo {author}
  {\bibfnamefont {J.}~\bibnamefont {{Helly}}}, \ and\ \bibinfo {author}
  {\bibfnamefont {C.}~\bibnamefont {{Frenk}}},\ }\href {\doibase
  10.1093/mnras/sts159} {\bibfield  {journal} {\bibinfo  {journal} {\mnras}\
  }\textbf {\bibinfo {volume} {428}},\ \bibinfo {pages} {1774} (\bibinfo {year}
  {2013})},\ \Eprint {http://arxiv.org/abs/1209.3018} {arXiv:1209.3018}
  \BibitemShut {NoStop}%
\bibitem [{\citenamefont {{Green}}\ \emph {et~al.}(2005)\citenamefont
  {{Green}}, \citenamefont {{Hofmann}},\ and\ \citenamefont
  {{Schwarz}}}]{GHS05}%
  \BibitemOpen
  \bibfield  {author} {\bibinfo {author} {\bibfnamefont {A.~M.}\ \bibnamefont
  {{Green}}}, \bibinfo {author} {\bibfnamefont {S.}~\bibnamefont {{Hofmann}}},
  \ and\ \bibinfo {author} {\bibfnamefont {D.~J.}\ \bibnamefont {{Schwarz}}},\
  }\href {\doibase 10.1088/1475-7516/2005/08/003} {\bibfield  {journal}
  {\bibinfo  {journal} {\jcap}\ }\textbf {\bibinfo {volume} {8}},\ \bibinfo
  {pages} {3} (\bibinfo {year} {2005})},\ \Eprint
  {http://arxiv.org/abs/arXiv:astro-ph/0503387} {arXiv:astro-ph/0503387}
  \BibitemShut {NoStop}%
\bibitem [{\citenamefont {{Boyarsky}}\ \emph {et~al.}(2009)\citenamefont
  {{Boyarsky}}, \citenamefont {{Lesgourgues}}, \citenamefont {{Ruchayskiy}},\
  and\ \citenamefont {{Viel}}}]{BLRV09}%
  \BibitemOpen
  \bibfield  {author} {\bibinfo {author} {\bibfnamefont {A.}~\bibnamefont
  {{Boyarsky}}}, \bibinfo {author} {\bibfnamefont {J.}~\bibnamefont
  {{Lesgourgues}}}, \bibinfo {author} {\bibfnamefont {O.}~\bibnamefont
  {{Ruchayskiy}}}, \ and\ \bibinfo {author} {\bibfnamefont {M.}~\bibnamefont
  {{Viel}}},\ }\href {\doibase 10.1088/1475-7516/2009/05/012} {\bibfield
  {journal} {\bibinfo  {journal} {\jcap}\ }\textbf {\bibinfo {volume} {5}},\
  \bibinfo {pages} {12} (\bibinfo {year} {2009})},\ \Eprint
  {http://arxiv.org/abs/0812.0010} {arXiv:0812.0010} \BibitemShut {NoStop}%
\bibitem [{\citenamefont {{Anderhalden}}\ and\ \citenamefont
  {{Diemand}}(2013)}]{AD12}%
  \BibitemOpen
  \bibfield  {author} {\bibinfo {author} {\bibfnamefont {D.}~\bibnamefont
  {{Anderhalden}}}\ and\ \bibinfo {author} {\bibfnamefont {J.}~\bibnamefont
  {{Diemand}}},\ }\href {\doibase 10.1088/1475-7516/2013/04/009} {\bibfield
  {journal} {\bibinfo  {journal} {\jcap}\ }\textbf {\bibinfo {volume} {4}},\
  \bibinfo {eid} {009} (\bibinfo {year} {2013})},\ \Eprint
  {http://arxiv.org/abs/1302.0003} {arXiv:1302.0003 [astro-ph.CO]} \BibitemShut
  {NoStop}%
\bibitem [{\citenamefont {{Ishiyama}}(2014)}]{Ishiyama14}%
  \BibitemOpen
  \bibfield  {author} {\bibinfo {author} {\bibfnamefont {T.}~\bibnamefont
  {{Ishiyama}}},\ }\href {\doibase 10.1088/0004-637X/788/1/27} {\bibfield
  {journal} {\bibinfo  {journal} {\apj}\ }\textbf {\bibinfo {volume} {788}},\
  \bibinfo {eid} {27} (\bibinfo {year} {2014})},\ \Eprint
  {http://arxiv.org/abs/1404.1650} {arXiv:1404.1650} \BibitemShut {NoStop}%
\bibitem [{\citenamefont {{The Fermi LAT
  Collaboration}}(2015)}]{2015JCAP...09..008T}%
  \BibitemOpen
  \bibfield  {author} {\bibinfo {author} {\bibnamefont {{The Fermi LAT
  Collaboration}}},\ }\href {\doibase 10.1088/1475-7516/2015/09/008} {\bibfield
   {journal} {\bibinfo  {journal} {Journal of Cosmology and Astro-Particle
  Physics}\ }\textbf {\bibinfo {volume} {2015}},\ \bibinfo {eid} {008}
  (\bibinfo {year} {2015})},\ \Eprint {http://arxiv.org/abs/1501.05464}
  {arXiv:1501.05464 [astro-ph.CO]} \BibitemShut {NoStop}%
\bibitem [{\citenamefont {Delos}\ \emph {et~al.}(2019)\citenamefont {Delos},
  \citenamefont {Bruff},\ and\ \citenamefont {Erickcek}}]{DBA19}%
  \BibitemOpen
  \bibfield  {author} {\bibinfo {author} {\bibfnamefont {M.~S.}\ \bibnamefont
  {Delos}}, \bibinfo {author} {\bibfnamefont {M.}~\bibnamefont {Bruff}}, \ and\
  \bibinfo {author} {\bibfnamefont {A.~L.}\ \bibnamefont {Erickcek}},\
  }\href@noop {} {\  (\bibinfo {year} {2019})},\ \Eprint
  {http://arxiv.org/abs/1905.05766} {arXiv:1905.05766 [astro-ph.CO]}
  \BibitemShut {NoStop}%
\bibitem [{\citenamefont {{Kawasaki}}\ \emph {et~al.}(1999)\citenamefont
  {{Kawasaki}}, \citenamefont {{Kohri}},\ and\ \citenamefont
  {{Sugiyama}}}]{KKS99}%
  \BibitemOpen
  \bibfield  {author} {\bibinfo {author} {\bibfnamefont {M.}~\bibnamefont
  {{Kawasaki}}}, \bibinfo {author} {\bibfnamefont {K.}~\bibnamefont {{Kohri}}},
  \ and\ \bibinfo {author} {\bibfnamefont {N.}~\bibnamefont {{Sugiyama}}},\
  }\href {\doibase 10.1103/PhysRevLett.82.4168} {\bibfield  {journal} {\bibinfo
   {journal} {Physical Review Letters}\ }\textbf {\bibinfo {volume} {82}},\
  \bibinfo {pages} {4168} (\bibinfo {year} {1999})},\ \Eprint
  {http://arxiv.org/abs/arXiv:astro-ph/9811437} {arXiv:astro-ph/9811437}
  \BibitemShut {NoStop}%
\bibitem [{\citenamefont {{Kawasaki}}\ \emph {et~al.}(2000)\citenamefont
  {{Kawasaki}}, \citenamefont {{Kohri}},\ and\ \citenamefont
  {{Sugiyama}}}]{KKS00}%
  \BibitemOpen
  \bibfield  {author} {\bibinfo {author} {\bibfnamefont {M.}~\bibnamefont
  {{Kawasaki}}}, \bibinfo {author} {\bibfnamefont {K.}~\bibnamefont {{Kohri}}},
  \ and\ \bibinfo {author} {\bibfnamefont {N.}~\bibnamefont {{Sugiyama}}},\
  }\href {\doibase 10.1103/PhysRevD.62.023506} {\bibfield  {journal} {\bibinfo
  {journal} {\prd}\ }\textbf {\bibinfo {volume} {62}},\ \bibinfo {pages}
  {023506} (\bibinfo {year} {2000})},\ \Eprint
  {http://arxiv.org/abs/arXiv:astro-ph/0002127} {arXiv:astro-ph/0002127}
  \BibitemShut {NoStop}%
\bibitem [{\citenamefont {{Hannestad}}(2004)}]{Han04}%
  \BibitemOpen
  \bibfield  {author} {\bibinfo {author} {\bibfnamefont {S.}~\bibnamefont
  {{Hannestad}}},\ }\href {\doibase 10.1103/PhysRevD.70.043506} {\bibfield
  {journal} {\bibinfo  {journal} {\prd}\ }\textbf {\bibinfo {volume} {70}},\
  \bibinfo {pages} {043506} (\bibinfo {year} {2004})},\ \Eprint
  {http://arxiv.org/abs/arXiv:astro-ph/0403291} {arXiv:astro-ph/0403291}
  \BibitemShut {NoStop}%
\bibitem [{\citenamefont {{Ichikawa}}\ \emph {et~al.}(2005)\citenamefont
  {{Ichikawa}}, \citenamefont {{Kawasaki}},\ and\ \citenamefont
  {{Takahashi}}}]{IKT05}%
  \BibitemOpen
  \bibfield  {author} {\bibinfo {author} {\bibfnamefont {K.}~\bibnamefont
  {{Ichikawa}}}, \bibinfo {author} {\bibfnamefont {M.}~\bibnamefont
  {{Kawasaki}}}, \ and\ \bibinfo {author} {\bibfnamefont {F.}~\bibnamefont
  {{Takahashi}}},\ }\href {\doibase 10.1103/PhysRevD.72.043522} {\bibfield
  {journal} {\bibinfo  {journal} {\prd}\ }\textbf {\bibinfo {volume} {72}},\
  \bibinfo {pages} {043522} (\bibinfo {year} {2005})},\ \Eprint
  {http://arxiv.org/abs/arXiv:astro-ph/0505395} {arXiv:astro-ph/0505395}
  \BibitemShut {NoStop}%
\bibitem [{\citenamefont {{Ichikawa}}\ \emph {et~al.}(2007)\citenamefont
  {{Ichikawa}}, \citenamefont {{Kawasaki}},\ and\ \citenamefont
  {{Takahashi}}}]{IKT07}%
  \BibitemOpen
  \bibfield  {author} {\bibinfo {author} {\bibfnamefont {K.}~\bibnamefont
  {{Ichikawa}}}, \bibinfo {author} {\bibfnamefont {M.}~\bibnamefont
  {{Kawasaki}}}, \ and\ \bibinfo {author} {\bibfnamefont {F.}~\bibnamefont
  {{Takahashi}}},\ }\href {\doibase 10.1088/1475-7516/2007/05/007} {\bibfield
  {journal} {\bibinfo  {journal} {\jcap}\ }\textbf {\bibinfo {volume} {5}},\
  \bibinfo {pages} {7} (\bibinfo {year} {2007})},\ \Eprint
  {http://arxiv.org/abs/arXiv:astro-ph/0611784} {arXiv:astro-ph/0611784}
  \BibitemShut {NoStop}%
\bibitem [{\citenamefont {{de Bernardis}}\ \emph {et~al.}(2008)\citenamefont
  {{de Bernardis}}, \citenamefont {{Pagano}},\ and\ \citenamefont
  {{Melchiorri}}}]{dBPM08}%
  \BibitemOpen
  \bibfield  {author} {\bibinfo {author} {\bibfnamefont {F.}~\bibnamefont {{de
  Bernardis}}}, \bibinfo {author} {\bibfnamefont {L.}~\bibnamefont {{Pagano}}},
  \ and\ \bibinfo {author} {\bibfnamefont {A.}~\bibnamefont {{Melchiorri}}},\
  }\href {\doibase 10.1016/j.astropartphys.2008.09.005} {\bibfield  {journal}
  {\bibinfo  {journal} {Astroparticle Physics}\ }\textbf {\bibinfo {volume}
  {30}},\ \bibinfo {pages} {192} (\bibinfo {year} {2008})}\BibitemShut
  {NoStop}%
\end{thebibliography}%

\begin{appendix}

\section{Gamma-Ray Constraints}
\label{gamma}

The angular distribution of the gamma rays produced through dark matter annihilating in microhalos is not the same as in the standard (smoothly distributed) scenario, but instead mimics the morphology predicted for the case of decaying dark matter particles. Over the range of dark matter models considered in this study, the strongest constraints are those derived from the Fermi Gamma-Ray Space Telescope's measurement of the high-latitude gamma-ray background. In evaluating the constraints on the dark matter's effective lifetime, we follow closely the approach described in Ref.~\cite{Blanco:2018esa}, altering only the branching fractions that are appropriate for the model under consideration and comparing the effective lifetime to the lower bound on the dark matter lifetime for particles with mass of $2 m_X$.

\begin{figure*}[t]
\includegraphics[scale=0.47]{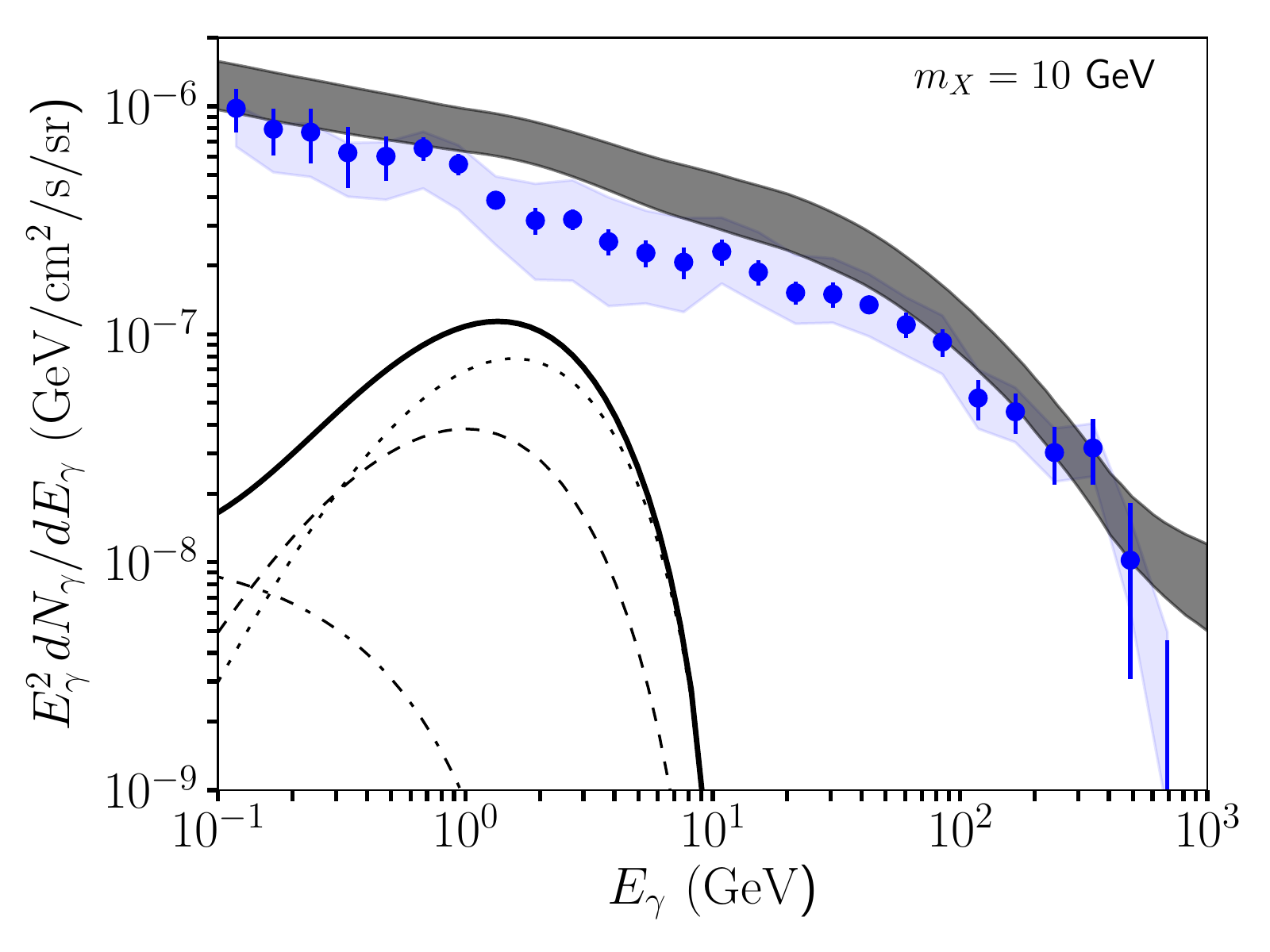} 
\includegraphics[scale=0.47]{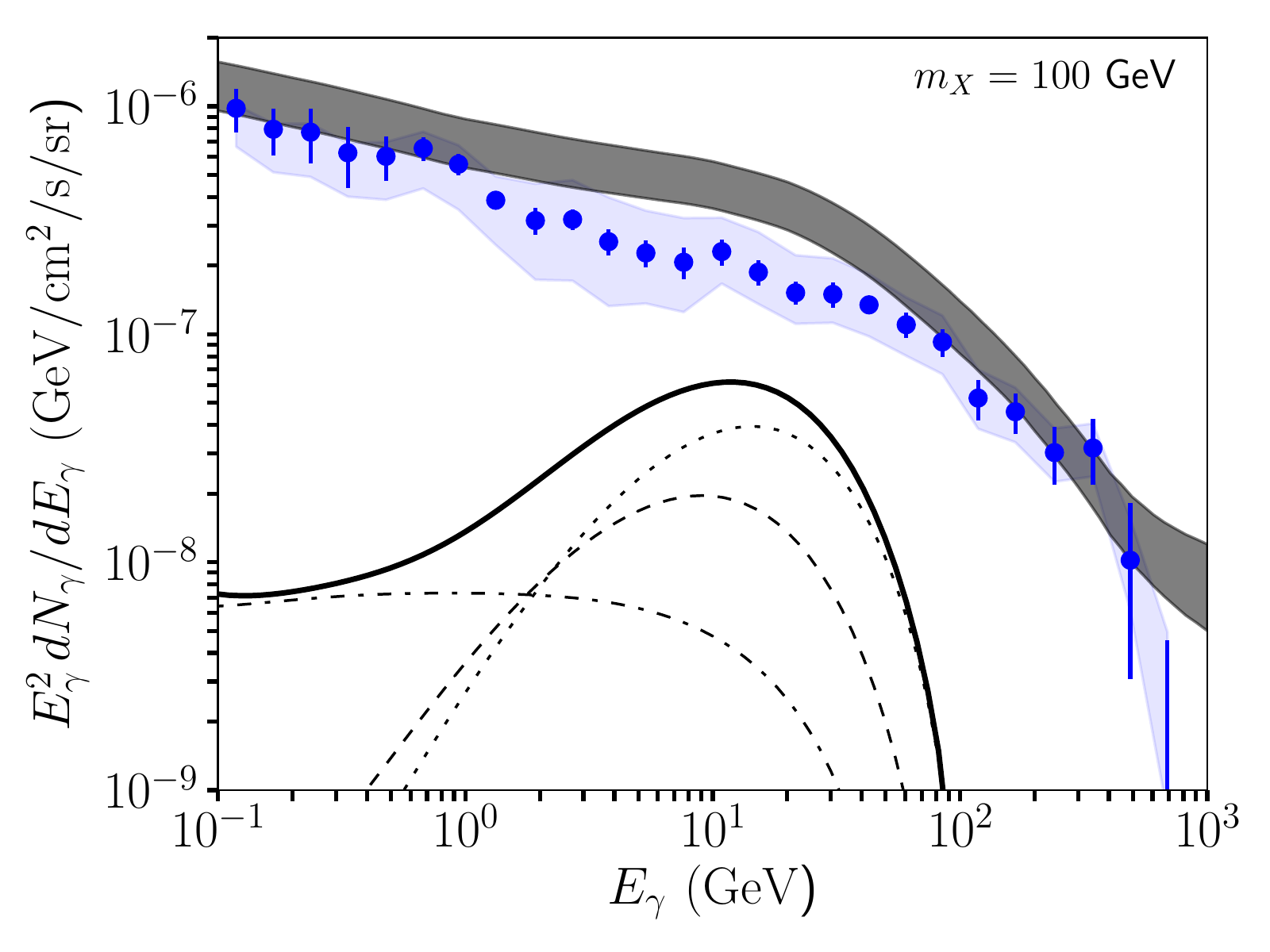} \\
\includegraphics[scale=0.47]{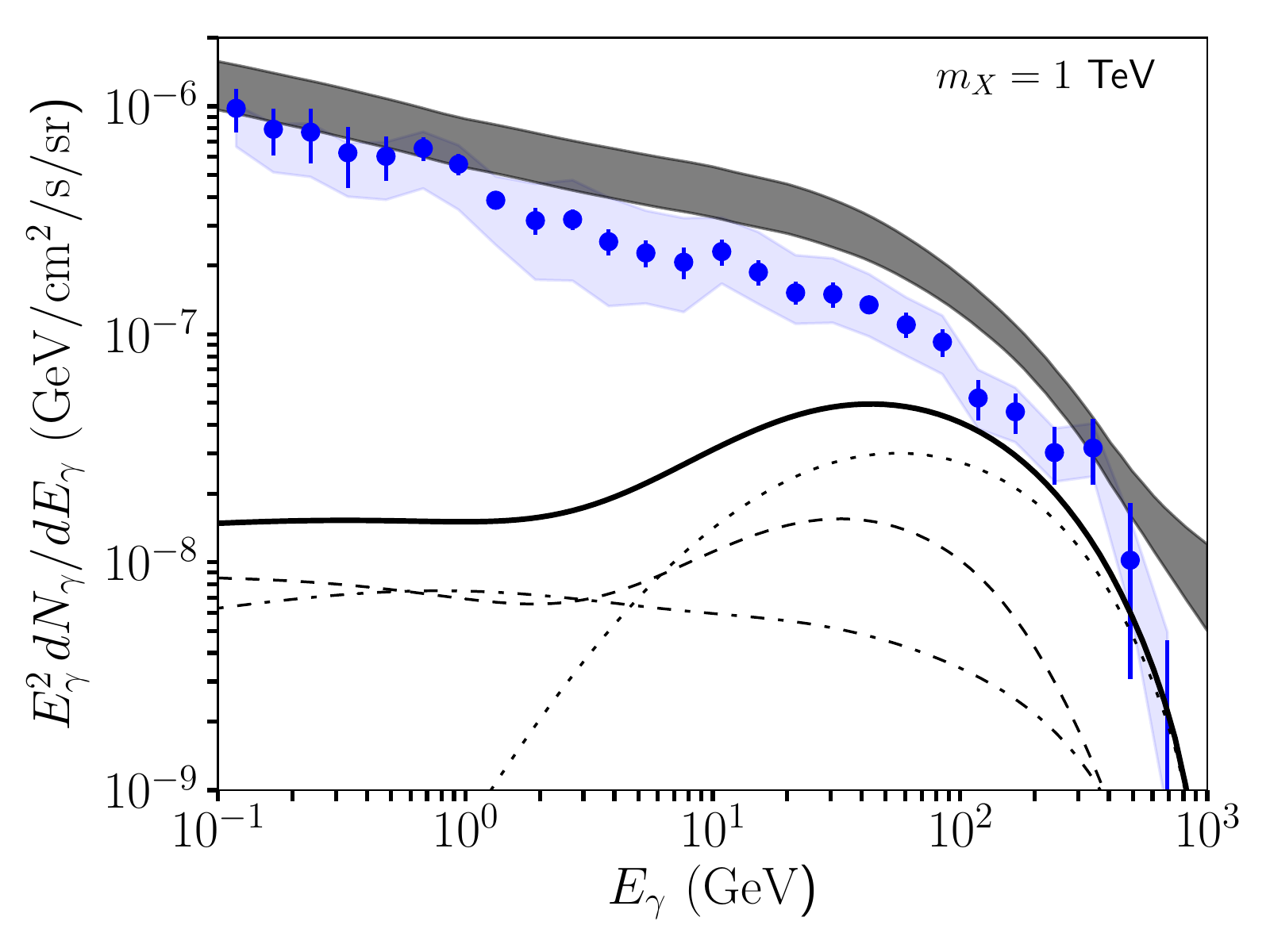} 
\includegraphics[scale=0.47]{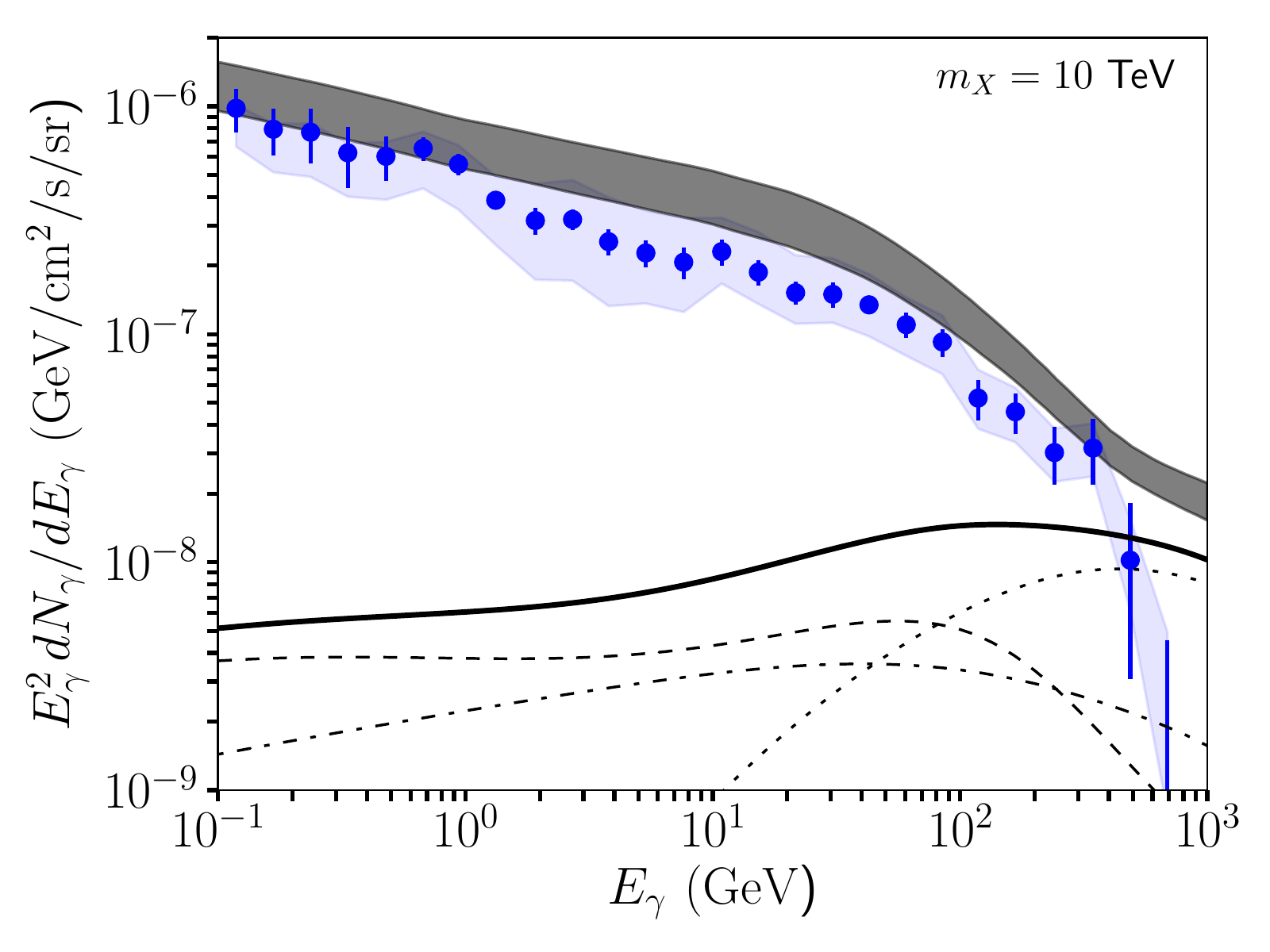} \\
\includegraphics[scale=0.47]{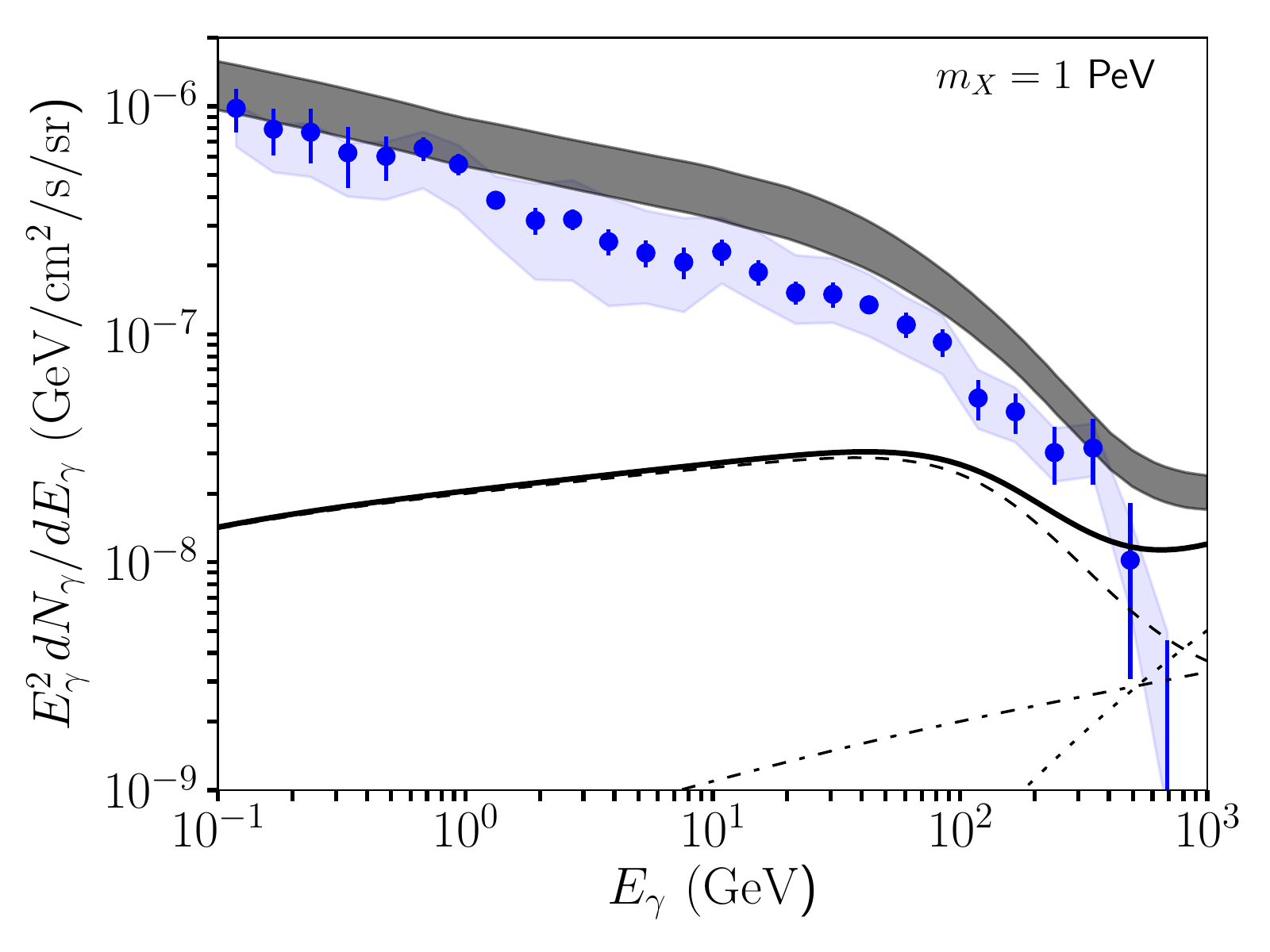} 
\caption{The contribution to the isotropic gamma-ray background from dark matter annihilations in the hidden sector model discussed in the text (with $m_X/m_{Z'}=20$). In each frame, the contributions from the Galactic Halo are shown as dotted and dot-dashed lines, representing the emission from direct production and from inverse Compton scattering, respectively. The dashed lines represent the cosmological contribution, including electromagnetic cascades. The solid lines denote the sum of these components. In each frame, the normalization (and corresponding dark matter annihilation rate) has been set to the maximum value allowed by the data (at the 95\% confidence level). For details, see Ref.~\cite{Blanco:2018esa}.}
\label{spec}
\vspace{0.5cm}
\end{figure*}

In Fig.~\ref{spec}, we plot the contribution to the isotropic gamma-ray background from dark matter annihilations in the hidden sector model under consideration for the case of $m_X/m_{Z'}=20$. In each frame, the overall normalization (and corresponding dark matter annihilation rate) has been set to the maximum value allowed by the data (at the 95\% confidence level), as evaluated following Ref.~\cite{Blanco:2018esa}.

\begin{figure}[ht]
\includegraphics[scale=0.47]{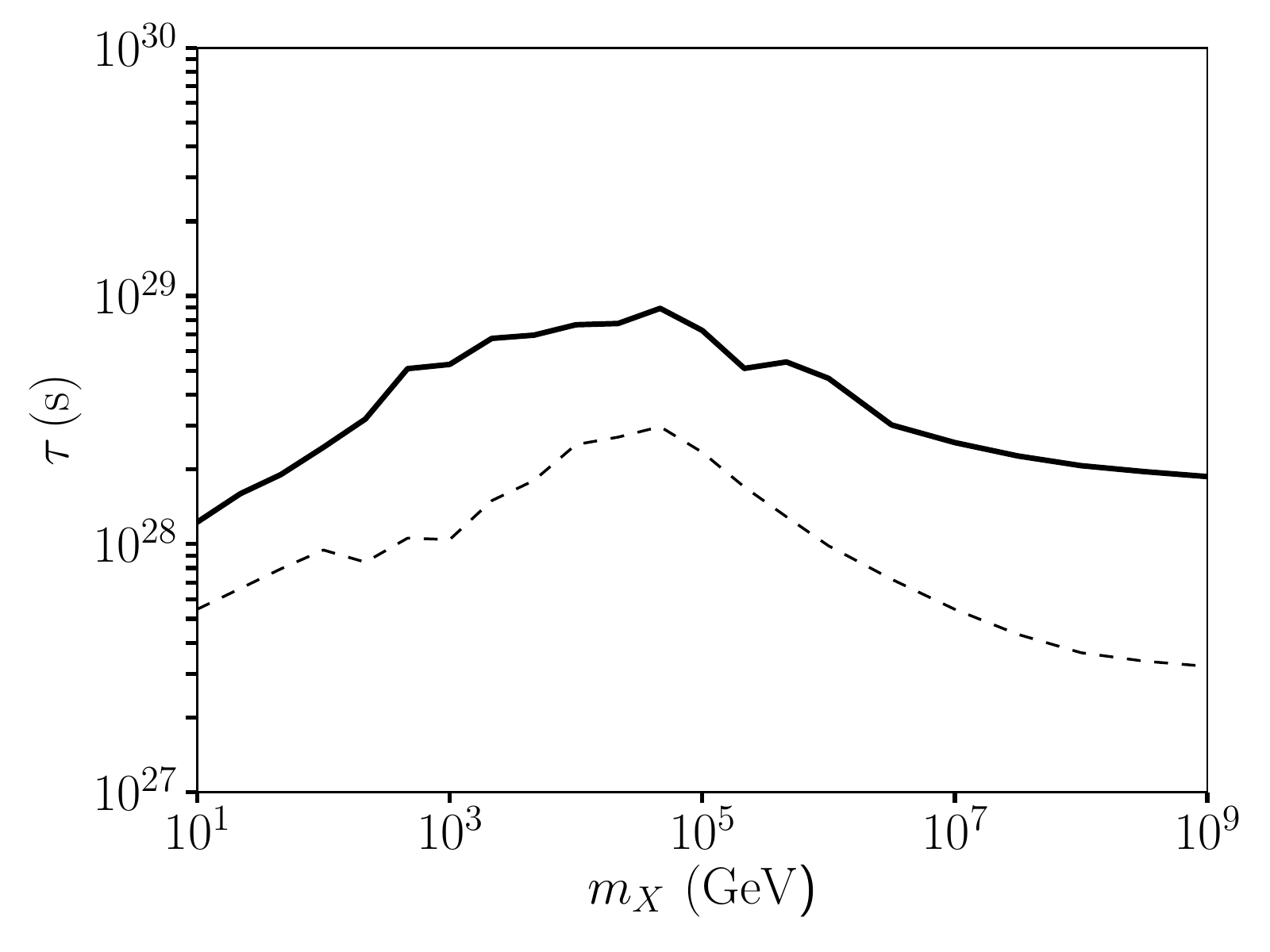} 
\caption{Lower limits on the dark matter's effective lifetime (95\% confidence level) as a function of mass. Here we have adopted $m_X/m_{Z'}=20$. The solid curves treat the systematic errors (shown as a blue band around the error bars in Fig.~\ref{spec}) as entirely independent and uncorrelated. At the other extreme, the dashed curve has been derived assuming that the systematic errors are fully correlated, moving upward or downward together in unison. These curves are the same as those shown with the same line types in Fig.~\ref{limits}. For details, see Ref.~\cite{Blanco:2018esa}.}
\label{limits2}
\end{figure}  

In Fig.~\ref{limits2}, we plot the limits on the effective lifetime of the dark matter particles in this model (again, for the case of $m_X/m_{Z'}=20$). Following Ref.~\cite{Blanco:2018esa}, the solid curves treat the systematic errors as entirely independent and uncorrelated. At the other extreme, the dashed curve has been derived assuming that the systematic errors are fully correlated, moving upward or downward together in unison.

\section{Threshold for Collapse During Radiation Domination}
\label{fitting}

During radiation domination, the linear collapse threshold is larger than its value during matter domination, $\delta_c > \delta_{c,0} \equiv 1.686$ .  The behavior of $\delta_c$ is a function of both the scale factor at collapse as well as the reheat temperature, as shown in Fig.~\ref{fig:deltac}.  In this appendix, we present a useful fitting formula for $\delta_c$.

\begin{figure}[t]
	\centering
	\includegraphics[width=\columnwidth]{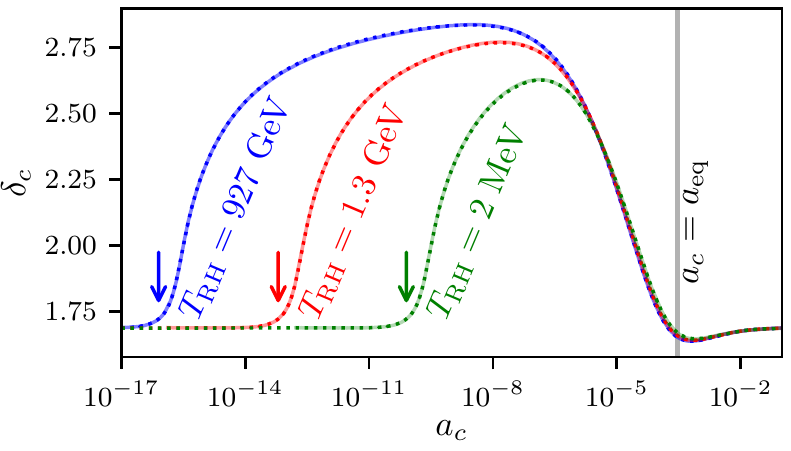}
	\caption{\label{fig:deltac} The threshold for linear collapse, $\delta_c$, as a function of the collapse scale factor, $a_c$, for three values of the reheat temperature, $T_\mathrm{RH}$. The light, solid lines represent $\delta_c$ computed using the spherical collapse model as described in Section~\ref{after}, while the dark, dotted lines employ the fitting formula presented in this Appendix.  The arrows mark the values of $a_\mathrm{RH}$ as defined in Eq.~(\ref{aRH}).}
\end{figure}

To begin, we define $a_\mathrm{RH}$ as the scale factor corresponding to temperature $T_\mathrm{RH}$ assuming radiation domination.  This definition implies that
\begin{equation}\label{aRH}
a_\mathrm{RH}=\left(\frac{3.91}{g_{*}(T_\mathrm{RH})}\right)^{1/3}\frac{T_0}{T_\mathrm{RH}},
\end{equation}
where $T_0 = \SI{2.3e-4}{eV}$ is the temperature today.  Note that this differs from other definitions of $a_\mathrm{RH}$ that appear in the literature, and we only use it as a parameter for our fitting formula. The following fitting formula for $\delta_\mathrm{c}$ is accurate to within 0.004 for $2\text{ MeV}< T_\mathrm{RH}< 1\text{ TeV}$ as long as $a_c\lesssim 0.1$ (dark energy is neglected) and ${a_c\gg T_0/T_\mathrm{dom}}$ (radiation prior to the EMDE is neglected):
\begin{equation}\label{deltac}
\delta_\mathrm{c}=
\delta_{\mathrm c,0}
+f\!\left(\frac{a_c}{a_\mathrm{RH}}\right)
g\!\left(\frac{a_c}{a_\mathrm{eq}},\ln\frac{a_\mathrm{RH}}{a_\mathrm{eq}}\right)
-h\!\left(\ln\frac{a_c}{a_\mathrm{eq}}\right).
\end{equation}
where $a_\mathrm{eq}=\Omega_\mathrm{r}/\Omega_\mathrm{m}$. The functions $f(x)$, $g(x)$ and $h(x)$ are defined as follows:
\begin{equation}
\begin{gathered}
f(x)=A\frac{\ln(1+x)-\frac{x+B x^2+C x^3}{1+D x+E x^2+F x^3}}{1+G (\ln(1+x)-2x/(2+x))},
\\
A = 0.3549,
\ \ 
B = -0.2331,
\ \ 
C = 0.0533,
\\ 
D = 0.4935,
\ \ 
E = -0.2092,
\\ 
F = 0.09327,
\ \ 
G = 0.2683,
\end{gathered}
\end{equation}
\begin{equation}
\begin{gathered}
g(x,y) = \left\{1+[A(y) x]^{B(y)}+[C(y) x]^{D(y)}\right\}^{-E(y)},
\\
A(y) = 135.2 \left(1+0.04734y+0.0006373y^2\right),
\\
B(y) = 1.093 \left(1+0.03256y+0.0005114y^2\right),
\\
C(y) = 17.87 \left(1+0.03501y+0.0003641y^2\right),
\\
D(y) = 3.187 \left(1+0.03283y+0.0005260y^2\right),
\\
E(y) = 0.05388 \left(1-0.7380y\right),
\end{gathered}
\end{equation}
and
\begin{equation}
\begin{gathered}
h(x)=A\mathrm{e}^{-B(x+C)^2},
\\
A = 0.07074,
\ \ 
B = 0.1180,
\ \ 
C = 0.4258.
\end{gathered}
\end{equation}

This fitting form was calibrated assuming no changes in the relativistic content of the universe, and incorporating them reduces $\delta_c$ by of order 1\%.  The difference is largest if reheating occurs while all Standard Model particles are still relativistic ($T_\mathrm{RH}\gsim 1$ TeV) and collapse occurs late in radiation domination ($a_c\sim 10^{-5}$), in which case $\delta_c$ is smaller by about 2\%.  Reducing $\delta_c$ increases the microhalo abundance and the resulting annihilation boost factor, but a $2\%$ reduction in $\delta_c$ increases $B_0$ by less than 20\%.

\end{appendix}

\end{document}